\documentclass[preprint,11pt,number]{elsarticle}
\usepackage[utf8]{inputenc}
\usepackage{amsmath}
\usepackage{amssymb}
\usepackage{amsthm}
\usepackage{graphicx}
\usepackage{listings}
\usepackage{appendix}
\usepackage{natbib}
\usepackage{array}
\usepackage{lipsum}
\usepackage{stmaryrd}
\usepackage{longtable}
\usepackage{multirow}
\usepackage{tikz}
\usepackage{caption}
\usepackage{subcaption}
\usepackage{lineno}
\usepackage{algorithm}
\usepackage[noend]{algpseudocode}
\usepackage{mathtools}
\usepackage{fullpage}
\usepackage{indentfirst}
\usepackage{xcolor}
\usepackage{courier} 
\usepackage{float}
\usepackage{hyperref}
\usepackage{caption} 
\usepackage{xcolor}
\usepackage{doi}
\usepackage{booktabs} 
\usepackage{orcidlink}
\usepackage{ulem}
\newcommand{\abs}[1]{\lvert #1 \rvert}

\newcommand{\cyr}[1]{\textcolor{black}{#1}}
\usetikzlibrary{shapes.geometric, arrows, positioning}

\tikzstyle{startstop} = [rectangle, rounded corners, minimum width=3.5cm, minimum height=1cm, text centered, draw=black, fill=red!60!yellow]
\tikzstyle{process} = [rectangle, minimum width=3.5cm, minimum height=1cm, text centered, draw=black, fill=blue!60!purple]
\tikzstyle{data} = [trapezium, trapezium left angle=70, trapezium right angle=110, minimum width=3.5cm, minimum height=1cm, text centered, draw=black, fill=orange!80!yellow]
\tikzstyle{arrow} = [thick,->,>=stealth]

\theoremstyle{definition}

\theoremstyle{remark}

\theoremstyle{plain}

\journal{Renewable Energy}
\begin{document}
\begin{frontmatter}

\title{\texttt{NICE}$^k$ Metrics: Unified and Multidimensional Framework for Evaluating Deterministic Solar Forecasting Accuracy}

\author[label1]{Cyril Voyant\corref{cor1}\orcidlink{0000-0003-0242-7377}}
\ead{cyril.voyant@mines-psl.eu} 
\cortext[cor1]{Corresponding Author}
\author[label2,label3]{Milan Despotovic\corref{cor1}\orcidlink{0000-0003-3144-5945}}
\ead{mdespotovic@kg.ac.rs} 
\author[label3]{Luis Garcia-Gutierrez\orcidlink{0000-0002-3480-1784}}
\ead{garcia_gl@univ-corse.fr} 
\author[label4,label1]{Rodrigo Amaro e Silva\orcidlink{0000-0002-2813-150X}}
\ead{rodrigo.amaro_e_silva@minesparis.psl.eu} 
\author[label5]{Philippe Lauret\orcidlink{0000-0003-2959-6146}}
\ead{philippe.lauret@univ-reunion.fr} 
\author[label6]{Ted Soubdhan\orcidlink{0000-0002-6402-8438}}
\ead{ted.soubdhan@univ-antilles.fr} 
\author[label7]{Nadjem Bailek\orcidlink{0000-0001-9051-8548}}
\ead{bailek.nadjem@univ-adrar.edu.dz} 
\address[label1]{Mines Paris, \texttt{PSL} University, Centre for Observation, Impacts, Energy (\texttt{O.I.E.}), 06904 Sophia Antipolis, France}
\address[label2]{Faculty of Engineering, University of Kragujevac, 6 Sestre Janjic, Kragujevac, Serbia}
\address[label3]{\texttt{SPE} Laboratory, \texttt{UMR CNRS} 6134, University of Corsica Pasquale Paoli, Ajaccio, France}
\address[label4]{Instituto Dom Luiz, Faculdade de Ciências, Universidade de Lisboa, Campo Grande, 1749-016 Lisboa, Portugal}
\address[label5]{\texttt{PIMENT} laboratory, University of La Réunion, 15, avenue René Cassin, 97715 Saint-Denis, France}
\address[label6]{\texttt{LARGE} Laboratory, University of Antilles, 97157 pointe à Pitre, France}
\address[label7]{Laboratory of Mathematics Modeling and Applications, Department of Mathematics and Computer Science - Faculty of Sciences and Technology, Ahmed Draia University of Adrar, Adrar, Algeria}

\date{\today}
\begin{abstract}
Accurate solar energy output prediction is fundamental to renewable energy source integration in electrical grids, keeping the system stable, and enhancing best-practice energy management. Nonetheless, conventional error measures, \textit{i.e.}, Root Mean Squared Error ($\mathtt{RMSE}$), Mean Absolute Error ($\mathtt{MAE}$), and Skill Scores ($\mathtt{SS}$), fail to represent solar irradiance forecasting's inherent multidimensional complexity. Among their known weaknesses, they lack sensitivity in relation to forecastability, depend on arbitrary baselines (\textit{e.g.}, clear-sky models), and have poor adaptability for operational requirements. In response to these challenges, this research introduces $\mathtt{NICE^k}$ measures (Normalized Informed Comparison of Errors with $k = 1, 2, 3, \Sigma$) as a novel evaluation framework designed to oﬀer a robust, ﬂexible, and multidimensional assessment of forecasting models. \cyr{Each $\mathtt{NICE^k}$ score corresponds to a different $\mathtt{L^k}$ norm: $\mathtt{NICE^1}$ emphasizes average errors, $\mathtt{NICE^2}$ is sensitive to large deviations, $\mathtt{NICE^3}$ amplifies outliers, and $\mathtt{NICE^{\Sigma}}$ aggregates their contributions.} The study employed synthetic \texttt{Monte Carlo} trials along with data from Spain \texttt{SIAR} network, which encompassed 68 meteorological stations in various climatic and geographic environments. The study focuses on predictive methods such as autoregressive models, Extreme Learning and smart persistence. Results showed that theoretical $\mathtt{NICE}$ metrics aligned perfectly with empirical values only when strict statistical assumptions held ($\mathtt{R}^2 \sim 1.0$ for $\mathtt{NICE^2}$), while composite metric $\mathtt{NICE^\Sigma}$ consistently outperformed conventional measures in discriminative power ($p<0.05$ vs $p>0.05$ for $\mathtt{nRMSE}$/$\mathtt{nMAE}$). Empirical inspection from increasing forecast horizons reveals $\mathtt{NICE^\Sigma}$ shows consistently significant \textit{p}-values (from $10^{-6}$ to 0.004), which outperform $\mathtt{nRMSE}$ and $\mathtt{nMAE}$ in achieving statistical significance. In addition, conventional measures such as $\mathtt{nRMSE}$, $\mathtt{nMAE}$, $\mathtt{nMBE}$, and $\mathtt{R}^2$ \cyr{cannot} discriminate among forecasting models in head-to-head comparisons since their \textit{p}-values lie above the 0.05 significance level. In contrast, $\mathtt{NICE}$ family shows greater discriminative strength in displaying $p < 0.001$, accompanied by wider and normally distributed values, which allows for greater inter-study comparability and generalizability. Theoretical and empirical validation using both synthetic and experimental solar irradiance datasets confirms the superior sensitivity and operational relevance of the $\mathtt{NICE^k}$ framework. These results highlight its potential to become a reliable benchmark for evaluating forecasting models in renewable energy applications.  \cyr{Finally, they support the adoption of $\mathtt{NICE^k}$ metrics as a unified, interpretable, and robust alternative to conventional error measures for deterministic solar forecasting evaluation.
}
\end{abstract}

\begin{graphicalabstract}
\begin{figure}[h!]
\centering
\resizebox{\textwidth}{!}{
\begin{tikzpicture}[node distance=2cm]
\node (input) [startstop,text=white] {Input Data: $\mathtt{I}$ (Solar Irradiance)};
\node (preprocess) [process, below of=input, text=white] {Step 1: Preprocessing (Forecasting with $\mathtt{X}$)};
\node (errorcalc) [process, below of=preprocess, text=white] {Step 2: Error Metrics ($\mathtt{RMSE}$, $\mathtt{MAE}$, $\mathtt{RMCE}$)};
\node (nicemetrics) [process, below of=errorcalc, text=white] {Step 3: $\mathtt{NICE}$ Metrics ($\mathtt{NICE}^1$, $\mathtt{NICE}^2$, $\mathtt{NICE}^3$, $\mathtt{NICE}^\Sigma$)};
\node (output) [startstop, below of=nicemetrics, text=white] {Output: Robust, Interpretable Results};
\draw [arrow] (input) -- (preprocess);
\draw [arrow] (preprocess) -- (errorcalc);
\draw [arrow] (errorcalc) -- (nicemetrics);
\draw [arrow] (nicemetrics) -- (output);
\node (baseline) [data, right of=errorcalc, xshift=7.9cm] {Comparison with Baseline (Persistence ; $\mathtt{P}$)};
\draw [arrow] (errorcalc.east) -- (baseline.west);
\node (dimensions) [data, left of=nicemetrics, xshift=-8cm] {Multidimensional Analysis ($\mathtt{L}^1$, $\mathtt{L}^2$, $\mathtt{L}^3$)};
\draw [arrow] (dimensions.east) -- (nicemetrics.west);
\node [below of=output, yshift=-1cm, text width=10cm, align=center, text=white] {\textbf{Key Advantages of $\mathtt{NICE}$ Metrics:}
\begin{itemize}
    \item Unified framework for $L^1$, $L^2$, and $L^3$ norms.
    \item Normalized and bounded ($0 \leq \mathtt{NICE}^k \leq 1$).
    \item Independent of arbitrary baselines.
    \item Sensitive to forecastability and variability.
\end{itemize}
};
\node [above of=baseline, yshift=0.5cm, text width=10cm, align=center, fill=yellow!20, draw=black] {
\textbf{Definition of $\mathtt{NICE}^2$ Metric:}
\begin{equation*}
\mathtt{NICE}^2 = \frac{\mathtt{RMSE_X}}{\mathtt{RMSE_P}}, \quad \mathtt{RMSE_P} = \sqrt{\frac{1}{n} \sum_{t=1}^n (y(t) - y(t-\Delta t))^2}
\end{equation*}
};
\node [above of=dimensions, yshift=0.5cm, text width=11cm, align=center, fill=blue!20, draw=black] {
\textbf{Generalization to $\mathtt{L^k}$:}
\begin{equation*}
\mathtt{NICE^k} = \frac{\mathtt{L^k}\mathtt{-Error_X}}{\mathtt{L^k}\mathtt{-Error_P}},  \quad \mathtt{L^k}\mathtt{-Error} = \left(\frac{1}{n} \sum_{t=1}^n |y(t) - \hat{y}(t)|^\mathtt{k} \right)^{1/\mathtt{k}}
\end{equation*}
};
\node [below of=baseline, yshift=-1cm, text width=10cm, align=center, fill=purple!20, draw=black] {
\textbf{Limitations of conventional Metrics:}
\begin{itemize}
    \item $\mathrm{RMSE}$ is scale-dependent.
    \item $\mathrm{MAE}$ lacks sensitivity to large deviations.
    \item Skill Scores depend on arbitrary baselines.
\end{itemize}
};
\end{tikzpicture}
}
\fcolorbox{black}{lightgray}{
    \parbox{0.95\textwidth}{\centering 
    \textbf{Overview of $\mathtt{NICE}$ Metrics Framework:} \\ 
    workflow, equations, and advantages for robust solar forecasting evaluation.}
}
\end{figure}
\end{graphicalabstract}

\begin{highlights}
\item New $\mathtt{NICE}$ Metrics framework for evaluating solar forecasting accuracy.
\item Addressing the limitations of conventional metrics like $\mathtt{RMSE}$ and $\mathtt{MAE}$ by ensuring boundedness and interpretability.
\item Two approaches are tested and the empirical one (versus theoretical) is the most insightful.
\item Robustness across multiple dimensions ($\mathtt{L}^1$, $\mathtt{L}^2$, $\mathtt{L}^3$ norms) and scenarios.
\item Validation of the framework using extensive solar irradiance datasets.
\end{highlights}

\begin{keyword} 
Renewable Energy Integration\sep Solar Forecasting \sep Energy Management Systems \sep \texttt{NICE} Metrics \sep Forecasting Accuracy \sep Multidimensional Metrics \sep Performance Assessment \sep Time Series Analysis
\end{keyword}

\end{frontmatter}
%
%

\section{Introduction}
Accurate forecasting of time series plays an important role in various scientific and engineering fields. From weather modeling to financial analytics, the ability to anticipate trends and deviations is the basis of modern decision-making \cite{lauret2012bayesianmodelcommitteeapproach}. In renewable energy, particularly solar energy forecasting, the need for precision is amplified by the stochastic nature of solar irradiance and the rapid integration of photovoltaics ($\mathtt{PV}$) into energy grids \cite{9279099}. These challenges underscore the necessity for robust evaluation frameworks to accurately assess and improve forecasting models \cite{manjili2018data,voyant2017machine}.
Deterministic forecasting methods, using Numerical Weather Prediction ($\mathtt{NWP}$) \cite{bacher2009short}, sky imagers \cite{yang2014solar}, and physics-based or statistical models \cite{taylor2016forecasting}, are widely utilized in solar forecasting. These methods aim to predict point values of solar irradiance or photovoltaic power output by minimizing deviations from observed data. Despite their widespread adoption, these approaches heavily depend on error metrics to evaluate model performance \cite{yang2020verification,zhang2015suite}. Unfortunately, these metrics often fail to address the intricate challenges of solar energy forecasting \cite{hansen2019arbiter}.

\subsection{Text Mining In Context} 
A structured text mining pipeline was applied to extract 70 publications (2014–2024) related to error metrics challenges in solar forecasting. Data were sourced from \texttt{Scopus} and enriched via \texttt{CrossRef} and \texttt{OpenAlex}. Filtering was conducted using sentence-transformer \texttt{MiniLM}, combining semantic similarity and keyword scoring. 
The corpus includes 39 journal articles, 20 conference papers, and 7 reviews. Open access rate is 34.3\%, with a mean citation count of 36.3. Saroha S. is the most prolific author (5 papers), while Sreeram V., and Mishra Y. lead in citations (829 each in average). Most frequent terms include \textit{numerical weather prediction}, \textit{solar forecasting}, and \textit{machine learning}. 
Yearly publication counts show a weak upward trend (\texttt{slope} = 0.65/year, \texttt{R}² = 0.26, \textit{p} = 0.13), suggesting topic stabilization. \texttt{China} (40 affiliations), \texttt{India} (31), and \texttt{Italy} (21) dominate geographically, contributing over 65\% of institutional records. 
Leading journals are \textit{Renewable Energy}, \textit{Energies}, and \textit{Solar Energy}; main publishers are \texttt{Elsevier} (22), \texttt{IEEE} (19), and \texttt{MDPI} (6). Despite the high volume of publications, most studies rely on standard error metrics such as $\mathtt{RMSE}$ and $\mathtt{MAE}$, with few explicitly introducing novel formulations, often implicitly embedded and thus difficult to extract via text mining. The analysis outlines a mature yet fragmented field, where methodological continuity often takes precedence over innovation.
This bibliometric snapshot underscores the central role of performance evaluation in solar forecasting research, while also revealing a persistent dependence on conventional statistical tools. To better situate this methodological landscape, the next section reviews the most commonly employed deterministic error metrics, their mathematical properties, and their limitations when applied to the stochastic nature of solar irradiance.

\subsection{Key Characteristics of Effective Solar Error Metrics}
To address the diverse requirements of solar forecasting, error metrics should exhibit the following properties \cite{vallance2017standardized}:
\begin{itemize}
    \item[\(\checkmark\)] \textbf{Linearity}: Metrics should proportionally reflect changes in model performance, ensuring transparency and ease of interpretation;
    \item[\(\checkmark\)] \textbf{Boundedness}: Metrics must operate within a defined range to enhance comparability;
    \item[\(\checkmark\)] \textbf{Horizon-invariance}: Metrics should remain applicable across different forecasting horizons, capturing both short-term and long-term dynamics;
    \item[\(\checkmark\)] \textbf{Statistical relevance}: Metrics should align with statistical principles, capturing key aspects such as mean deviations, variances, and higher-order moments when necessary;
    \item[\(\checkmark\)] \textbf{Independence from external models}: Metrics should not rely on assumptions or outputs from ancillary models, ensuring universal applicability;
    \item[\(\checkmark\)] \textbf{Applicability to energy systems}: Metrics should reflect operational impacts, such as ramping errors or energy yield discrepancies \cite{marquez2013proposed};
    \item[\(\checkmark\)] \textbf{Sensitivity to extremes}: Metrics must adequately weight outliers due to the high operational risks associated with large forecasting errors.
    \item[\(\checkmark\)] \textbf{Scale invariance}: Metrics should be dimensionless and independent of the magnitude or seasonal range of the predicted variable, ensuring fair comparisons across different time periods, locations, or forecasting conditions.
\end{itemize}

\subsection{Challenges with Common Forecasting Metrics in Solar Energy}
Deterministic solar forecasts are most commonly evaluated using a relatively small set of error metrics \cite{yang2020verification}, lauded for their simplicity and ease of computation \cite{willmott2005advantages}.
These include the Root Mean Squared Error ($\mathtt{RMSE}$), Mean Absolute Error ($\mathtt{MAE}$), Mean Bias Error ($\mathtt{MBE}$), Skill Scores ($\mathtt{SS}$), and the coefficient of determination ($\mathtt{R^2}$). Normalized versions such as the normalized $\mathtt{RMSE}$ ($\mathtt{nRMSE}$) are frequently used to enable scale-independent comparisons across datasets and forecast horizons. Some normalize by the mean, others by the maximum value, the range (max - min), or the root mean square.
Each of these metrics offers a distinct analytical perspective. $\mathtt{RMSE}$ places greater weight on large deviations, making it sensitive to infrequent but significant forecast errors. While this property can be desirable when such events are operationally critical, it can also penalize models that perform reliably under normal conditions. $\mathtt{MAE}$, on the other hand, provides a more balanced measure of average forecast deviation, being less influenced by outliers, but it may obscure the impact of extreme values that are often crucial in grid operation and planning. $\mathtt{MBE}$ serves to quantify systematic bias by evaluating the mean signed error, yet it conveys no information about the spread or distribution of the residuals. 
$\mathtt{R^2}$ provide not only normalized but also bounded performance indicators, estimating the proportion of observed variance explained by the forecast, offering a high-level indicator of model fit. However, it is insensitive to systematic errors and largely unresponsive to errors in the time domain, which can obscure temporal misalignments \cite{marquez2013proposed}. In contrast, $\mathtt{SS}$ benchmarks model performance relative to a specified baseline and metric to a given metric. For example, an earlier study by Murphy focuses on  meteorological variables in general,using mean squared error ($\mathtt{MSE}$) as the metric and climatology as the baseline \cite{SS}. A more recent work by Marquez and Coimbra explores this concept specifically to solar forecasting, proposing a derivation which is said to be equivalent to a $\mathtt{SS}$ based on $\mathtt{RMSE}$ and smart persistence, considering either the clearness or clear-sky index \cite{marquez2013proposed}. While such benchmarking adds interpretive value and is  horizon-invariant, it remains confined to the dimensions captured by the selected metric. Moreover, the use of different baselines in the literature (\textit{i.e.}, lack of standardization) complicates the interpretation and comparability of results. In particular, using smart persistence as a baseline introduces additional variability, as its performance is influenced by the chosen reference model. For example, when it is based on the clear-sky index, there is a wide range of clear-sky models that differ in complexity and accuracy, depending on factos such as geographic location, weather conditions, timestamp, and underlying modeling assumptions \cite{perez2017value,lorenz2009forecast}. 
However, their application to solar forecasting exposes critical shortcomings due to the non-stationary and intermittent nature of solar irradiance, which violates the assumptions of time-invariant error distributions. This leads to biased skill assessments, as benchmark performance can vary significantly with forecasting horizon, weather regimes, and seasonal cycles.
While statistical error metrics are key indicators in solar forecasting, the very definition of forecast quality remains context-dependent \cite{perez2017value}. Across fiels such as meteorology, finance, and energy, it is recognized that forecast quality cannot be reduced solely to mathematical metrics like $\mathtt{RMSE}$. For example, in agro-meteorology, quality might be evaluated based on water savings rather than statistical accuracy. This study adopts a mathematical perspective, assessing forecast quality primarily as a statiscal indicator, rather than a user-centric measure. Forecasting requirements vary widely: meteorologists employ probabilistic scores (\textit{e.g.}, $\mathtt{Brier Score}$, $\mathtt{CRPS}$) \cite{YANG2019410}, while financial analysts use risk-adjusted metrics (\textit{e.g.}, $\mathtt{VaR}$, $\mathtt{Sharpe Ratio}$). Unlike humidity or temperature forecasting, solar radiation prediction is particularly challenging due to its intermittency and cloud-dependent variability, reinforcing the need for adapted evaluation metrics. \citep{vallance2017standardized} highlight the limitations of $\mathtt{nRMSE}$ and proposed a promising bounded alternative based on temporal distortion. While interesting, this approach may be somewhat complex to implement in practice.

\subsection{Objectives of This Study}
The diversity of metrics in solar forecasting reflects the multifaceted nature of forecasting performance and the inherent trade-offs involved \cite{zhang2015suite}. For instance, a model that excels in $\mathtt{RMSE}$ may underperform in $\mathtt{MAE}$ due to its sensitivity to large errors \cite{foley2012current}. Moreover,in a simple case using irradiance data from Ajaccio (France) over 90 winter days, $\mathtt{nRMSE}$ (normalized using irradiance average) values ranged from $0.047$ to $0.288$ depending on the temporal resolution (5-minute to daily) with a one-step persistence model. Such variability highlights that $\mathtt{nRMSE}$ alone cannot reliably reflect forecast quality, as it is sensitive to the forecasting horizon, data resolution, and normalization method, all of which vary widely across studies. These discrepancies introduce ambiguity in model evaluation, leading to metric degeneracy, where no single metric sufficiently differentiates models across all performance aspects. This undermines decision-making, particularly when a consistent and transparent evaluation framework is required \cite{Singla2021}. Furthermore, the lack of consensus on standard metrics prevents meaningful model ranking and inhibits cross-study comparisons, ultimately slowing progress in the field.

These challenges underscore the need for a more comprehensive and standardized evaluation methodology that can accommodate the complexity of forecasting tasks while preserving interpretability and comparability. To this end, the present study proposes a novel framework designed to address these limitations and provide a more robust foundation for model assessment. The framework ensures that metrics are bounded, operating within a defined range to enable consistent and meaningful comparisons across models and datasets. It incorporates a multi-dimensional evaluation approach, designed to balance sensitivity to typical errors, large deviations, and extreme outliers, thereby capturing the full spectrum of forecasting performance. Additionally, the framework emphasizes the normalization process, contextualizing performance relative to the inherent predictability of the data \cite{10.1063/5.0042710}. This approach enables objective and reliable comparisons across diverse conditions and operational scenarios. By integrating statistical, operational, and engineering perspectives, the proposed methodology provides a unified, transparent, and robust framework for evaluating solar forecasting models. Its design ensures broad applicability across various datasets, forecast horizons, and energy systems, fostering more consistent and interpretable performance assessments in the field.

\cyr{The specific contribution of this study is the introduction of the $\mathtt{NICE^k}$ metric family, a bounded, data-driven framework for multidimensional evaluation of deterministic forecasts, which resolves key limitations of $\mathtt{nRMSE, MAE}$, and skill scores. Its formulation allows both theoretical interpretability and empirical robustness across datasets and forecast horizons. Beyond model comparison, the proposed approach is intended to support operational decision-making in energy systems, including forecasting pipeline optimization, model selection, and grid-aware forecast evaluation.
}

%
%

\section{Methods}
\label{sec:methodology}
Conventional error measures such as $\mathtt{RMSE}$ and $\mathtt{MAE}$ are widely used due to their simplicity and recognized relevance. However, their limitations, such as sensitivity to scale, lack of standardization and inability to account for variability, underline the need for improved measures adapted to the challenges of solar forecasting. To address these challenges, a new process is detailed in this section. Since the $\mathtt{RMSE}$, widely used in solar forecasting \cite{zhang2015suite, marquez2013proposed}, relies on the $\mathtt{L}^2$ norm, known for its desirable mathematical properties (\textit{e.g.}, continuity and differentiability), the analysis will begin within the $\mathtt{L}^2$ error framework. The pioneering work of Ricardo Marquez and Carlos Coimbra in formalizing the $\mathtt{RMSE}$-based Forecast Skill \cite{marquez2013metric} metric within the context of solar forecasting serves as a key inspiration for the present study. Interestingly, according to conclusions of meta-analyses such as \cite{ttt,Nguyen2022AMO}, our proposed metric shows strong similarity to previously observed ones, but under certain integrated conditions. It  differs in important ways, most notably through its formulation within an $\mathtt{L}$-norm framework and its intrinsic normalization between 0 and 1. Unlike the classical metrics, which often depends on physical or knowledge-based reference models (\textit{i.e.} clear-sky model for clear-sky index and top of the atmosphere irradiance for clearness index), our approach remains fully data-driven and norm-consistent, aiming to offer a more general and theoretically grounded evaluation of forecast performance.

\subsection{Normalized Informed Comparison of \texttt{$L^2$}-Errors (\textit{i.e.} \texttt{NICE\ensuremath{^2}})}
$\mathtt{RMSE}$ quantifies the average magnitude of the errors between predicted and actual values. It is defined as \cite{zhang2015suite}:
    \begin{equation}
    \mathtt{RMSE} = \sqrt{\frac{1}{n} \sum_{t=1}^n (y(t) - \hat{y}(t))^2},
    \label{eq:rmse}
    \end{equation}
where $y(t)$ is the true value at time $t$, $\hat{y}(t)$ is the predicted value at time $t$ and $n$ is the length of the time series (or the evaluation period).
While $\mathtt{RMSE}$ is a useful metric \cite{marquez2013metric}, its scale depends on the magnitude of the data. This makes direct comparisons of $\mathtt{RMSE}$ values across different time series or datasets with varying scales challenging. To enable meaningful comparisons and provide a relative measure of model performance, a normalized metric is introduced: the Normalized Root Mean Squared Error for model $\mathtt{X}$, denoted as $\mathtt{nRMSE_X}$. While this normalization does not produce bounded values, an improved metrics called $\mathtt{NICE^2}$ is proposed in this study. This metric allows the forecasting model to be evaluated relative to a simple persistence (or naive; $\mathtt{P}$), where the forecast at time $t$ is simply the observed value at previous time $t-\Delta t$, (\textit{i.e.}, $\hat{y}(t) = y(t-\Delta t)$ where $\Delta t$ is the time step). $\mathtt{NICE^2}$ is defined from $\mathtt{RMSE_P}$, the $\mathtt{RMSE}$ of the persistence model, which is computed as:
    \begin{equation}
    \mathtt{RMSE_P} = \sqrt{\frac{1}{n} \sum_{t=1}^n (y(t) - y(t-\Delta t))^2}.
    \label{eq:rmse_persistence}
    \end{equation}
To define the normalized metric, we start from the general form:
\begin{equation}
\mathtt{NICE^2} = \alpha \mathtt{RMSE_X},
\label{eq:general_normalization}
\end{equation}
where $\alpha$ is a normalization factor computed to satisfy the following conditions:
\begin{enumerate}
    \item[\(\checkmark\)] Perfect prediction ($y(t) = \hat{y}(t)$, $\forall t$) should yield $\mathtt{NICE^2}=0$;
    \item[\(\checkmark\)] Worst-case prediction (\textit{i.e.}, persistence with $\hat{y}(t) = y(t-\Delta t)$) should yield $\mathtt{NICE^2}=1$.
\end{enumerate}
From the general form of $\mathtt{NICE^2}$ (Equation \ref{eq:general_normalization}), these conditions lead to:
\begin{enumerate}
    \item[\(\checkmark\)] If $\mathtt{RMSE_X} = 0$ (perfect prediction), then $\mathtt{NICE^2} = \alpha \times 0 = 0$;
    \item[\(\checkmark\)] If $\mathtt{RMSE_X} = \mathtt{RMSE_P}$ (persistence benchmark), then  
    \begin{equation*}
    \mathtt{NICE^2} = \alpha \times \mathtt{RMSE_P} = 1, \quad \text{which implies} \quad \alpha = \frac{1}{\mathtt{RMSE_P}}.
    \end{equation*}
\end{enumerate}
Substituting this value of $\alpha$ into Equation \eqref{eq:general_normalization}, the final expression is obtained:
\begin{equation}
\mathtt{NICE^2} = \frac{\mathtt{RMSE_X}}{\mathtt{RMSE_P}}.
\label{eq:derived_nrmse}
\end{equation}
This derivation naturally leads to the standard definition of $\mathtt{NICE^2}$, confirming that it emerges as a well-normalized metric that satisfies the desired properties: a lower bound at 0 for perfect predictions and an upper bound at 1 for persistence. This ensures its interpretability and robustness as a comparative performance measure in solar forecasting. 
For practical models, the values of $\mathtt{NICE^2}$ range from 0 to 1, assuming that the worst reasonable benchmark is the persistence model. While it is theoretically possible to construct an even worse model, in practice (especially for time steps beyond a few minutes), if a model performs worse than simple persistence, its usefulness is highly questionable, and little insight can be gained from its evaluation.

\subsection{\texttt{$L^k$} generalization : Empirical Analysis of \texttt{NICE\ensuremath{^k}} Metrics}
\label{empirical}
The $\mathtt{NICE^k}$ framework extends normalized error metrics to $\mathtt{L^k}$ norms, where $\mathtt{k}$ controls the sensitivity of the metric to different magnitudes of forecasting errors. The generalized $\mathtt{NICE^k}$ metric for a model $\mathtt{X}$ is defined as:
\begin{equation}
    \mathtt{NICE^k} = \frac{\mathtt{L^k}\mathtt{-Error_X}}{\mathtt{L^k}\mathtt{-Error_P}},
\end{equation}
where $\mathtt{L^k-Error_X}$ is the $\mathtt{L^k}$-norm error of model $\mathtt{X}$, and $\mathtt{L^k-Error_P}$ is the corresponding error for the persistence model $\mathtt{P}$. The $\mathtt{L^k}$-norm error is computed as:
\begin{equation}
    \mathtt{L^k\mathtt{-Error}} = \left( \frac{1}{n} \sum_{t=1}^n |y(t) - \hat{y}(t)|^\mathtt{k} \right)^{\frac{1}{\mathtt{k}}}.
\end{equation}
The choice of $\mathtt{k}$ influences how the metric responds to forecasting errors. For $\mathtt{k} = 1$, the metric reduces to the normalized Mean Absolute Error ($\mathtt{MAE}$), which treats all errors equally and is robust to outliers. When $\mathtt{k} = 2$, it corresponds to the normalized Root Mean Squared Error ($\mathtt{RMSE}$), which emphasizes larger deviations due to the squaring effect. For $\mathtt{k} = 3$, the metric further amplifies large errors, making it particularly relevant for applications where extreme deviations are critical, such as energy grid management or extreme event forecasting. This defines the root mean cubic error ($\mathtt{RMCE}$) metrics.
While $\mathtt{NICE^3}$ may be useful in such high-risk scenarios, increasing $\mathtt{k}$ beyond 3 generally reduces the interpretability of the metric. For large $\mathtt{k}$, the metric becomes overly dominated by extreme errors, making it less relevant for practical forecasting, particularly for time series like solar radiation, where variability increases unpredictability.
For short-term forecasting, $\mathtt{NICE^1}$ and $\mathtt{NICE^2}$ are generally sufficient. $\mathtt{NICE^1}$ provides robustness by reducing sensitivity to extreme errors, while $\mathtt{NICE^2}$ balances penalizing large deviations while preserving overall accuracy. Although $\mathtt{NICE^3}$ can add value in specialized cases, it is less commonly applicable.
In practice, the choice of $\mathtt{k}$ should be guided by the application context. $\mathtt{NICE^2}$ remains a standard choice due to its balance between error magnitude sensitivity and interpretability. $\mathtt{NICE^1}$ is preferable when robustness against outliers is essential. However, for $\mathtt{k} > 3$, the metric loses general usefulness as extreme errors dominate, limiting its practical relevance for forecasting tasks. Some important properties of $\mathtt{NICE^k}$ are demonstrated in \ref{annex:properties} (invariant to scale and translation transforms).

\subsection{Link with the Autocovariance: Theoretical Statement}
Let $\{y(t)\}$ be a weakly stationary time series. The lag $n \in \mathbb{N}_+$ corresponds to a temporal shift between observations, such that $y(t - n\Delta t)$ represents the value of the process at time $t - n\Delta t$, with $\Delta t$ denoting the uniform time step. When $n  = 1$, we consider a lag-1 difference, comparing consecutive values $y(t)$ and $y(t-\Delta t)$. A common approach to quantify the variability of a time series over time is to analyze the expected squared difference between values separated by $n $ time steps. Specifically, for lag-1, this is given by:
\begin{equation}
    \mathbb{E}[(y(t) - y(t-\Delta t))^2] = 2 \sigma^2 (1 - \rho(1)),
    \label{eq:stationary_squared_diff}
\end{equation}
where $\rho(n )$ is the lag-$n $ autocorrelation coefficient, defined from the covariance function $C(n)$ as $\rho(n) = \frac{C(n )}{\sigma^2}$.  
This result relies on the assumption of weak stationarity, which requires that the mean remains constant over time and that the variance is time-invariant, \textit{i.e.}, $\sigma^2(y(t)) = \sigma^2(y(t-n\Delta t)) = \sigma^2$.  
Although full weak stationarity may not hold for all time series, many practical signals (such as global horizontal irradiance: $\mathtt{I}$ in solar energy) exhibit approximate stationarity at small lags ($n =1$). This approximation is reasonable due to strong short-term autocorrelation induced by local meteorological conditions. In particular, the assumption is generally valid in two key time-scale ranges:
\begin{itemize}
    \item[\(\checkmark\)] High-frequency data ($\in [0, 1]$ hour): Intra-hour variations in $\mathtt{I}$ are largely driven by short-term fluctuations such as cloud movement, maintaining a relatively stable variance over short intervals.  
    \item[\(\checkmark\)] Daily-scale data ($\in [24$ hours, 1 month]): When aggregated at a daily resolution, $\mathtt{I}$ exhibits consistent statistical properties over several weeks, except during strong seasonal transitions.  
\end{itemize}
However, if the time step between observations is too large (\textit{e.g.}, monthly data), the assumption $\sigma^2(y(t)) = \sigma^2(y(t-\Delta t))$ may no longer hold due to long-term trends, seasonal effects, or structural changes in the series.  
The persistence model, defined as $\hat{y}(t) = y(t-\Delta t)$, has a root mean squared error given by \eqref{eq:rmse_persistence}. Substituting Equation \eqref{eq:stationary_squared_diff} into this equation (under the stationarity assumption) yields:
\begin{equation}
    \mathtt{RMSE_P} \approx \sqrt{2 \sigma^2 (1 - \rho(1))}.
    \label{eq:persistence_rmse_autocorrelation}
\end{equation}
$\mathtt{NICE^2}$ is then defined as:
\begin{equation}
    \mathtt{NICE^2} = \frac{\mathtt{RMSE_X}}{\mathtt{RMSE_P}} \approx \frac{\sqrt{\frac{1}{n} \sum_{t=1}^n (y(t) - \hat{y}(t))^2}}{\sqrt{2 \sigma^2 (1 - \rho(1))}} \approx \frac{1}{\sigma \sqrt{2} } \frac{\sqrt{\frac{1}{n} \sum_{t=1}^n (y(t) - \hat{y}(t))^2}}{\sqrt{1 - \rho(1)}}.
    \label{eq:general_normalization2}
\end{equation}
This formulation establishes a direct connection between $\mathtt{NICE^2}$ and the temporal dependency structure of the time series ($\rho(1)$). It also highlights why the persistence model serves as a natural reference for normalization, especially for series exhibiting strong short-term autocorrelation.
A similar derivation can be applied to define $\mathtt{NICE^1}$ and $\mathtt{NICE^3}$, leading to:
\begin{equation}
    \mathtt{NICE^1} \approx \frac{\mathtt{MAE_X}}{\sqrt{\frac{2}{\pi}} \sqrt{2 \sigma^2 (1 - \rho(1))}} \quad \text{and} \quad
    \mathtt{NICE^3} \approx \frac{\mathtt{RMCE_X}}{\left(\frac{4 \sqrt{2}}{\sqrt{\pi}}   (2 \sigma^2 (1 - \rho(1)))^{3/2} \right)^{1/3}}.
\end{equation}
Demonstrations are available in \ref{NICE1&2}. This generalization extends the framework to different error norms, further reinforcing the robustness of the $\mathtt{NICE^k}$ family in assessing forecasting performance while incorporating autocovariance properties. However, different values of $\mathtt{k}$ introduce distinct theoretical constraints. While a stationarity assumption is required for $\mathtt{k}=2$, an additional normality assumption is necessary for $\mathtt{k}=1$ and $\mathtt{k}=3$ to compute the expectation values $\mathbb{E}[\abs{y(t) - y(t-\Delta t)}]$ and $\mathbb{E}[(y(t) - y(t-\Delta t))^3]$ (see \ref{annex:limitation}).

\subsection{Extension to Forecasting Horizons $t + n\Delta t$}
When extending the forecasting horizon to $t + n\Delta t $, where $n$ is the lag horizon and $\Delta t$ is the time step, the $\mathtt{RMSE_{P}}$ can be generalized to reflect the performance of a model over this horizon. The persistence model, in this case, predicts $\hat{y}(t+n\Delta t) = y(t)$, assuming no change in the observed variable. At horizon $n$, the $\mathtt{RMSE}$ of the persistence model is given by:
\begin{equation}
    \mathtt{RMSE_{P}} = \sqrt{\frac{1}{n} \sum_{t=1}^n (y(t+n\Delta t) - y(t))^2}.
\end{equation}
Under the assumption of weak stationarity, the error of the persistence model is directly related to the lag-$n$ autocorrelation coefficient $\rho(n)$. Specifically, the expected squared difference between $y(t+n\Delta t)$ and $y(t)$ is given by:
\begin{equation}
    \mathbb{E}[(y(t+n\Delta t) - y(t))^2] = 2 \sigma^2 (1 - \rho(n)).
\end{equation}
Here, $\rho(n)$ represents the autocorrelation at lag $n$, defined as:
\begin{equation}
    \rho(n) = \frac{C(t+n\Delta t, t)}{\sigma^2}.
\end{equation}
For short horizons (\textit{i.e.}, when $n\Delta t$ remains within a few hours), the assumption of weak stationarity is generally valid for many practical time series, such as $\mathtt{I}$. This is because, at these scales, variability is primarily driven by predictable meteorological patterns (\textit{e.g.}, cloud motion, diurnal cycle), ensuring a relatively stable variance and autocorrelation structure.
Using the link between the persistence error and $\rho(n)$, $\mathtt{NICE^2}$ at horizon $n$ becomes:
\begin{equation}
    \mathtt{NICE^2} = \frac{\sqrt{\frac{1}{n} \sum_{t=1}^n (y_{t} - \hat{y}_{t})^2}}{\sqrt{2  \sigma^2 (1 - \rho(n))}}=\frac{1}{\sigma \sqrt2 }\frac{\sqrt{\frac{1}{n} \sum_{t=1}^n     (y_{t} - \hat{y}_{t})^2}}{\sqrt{1 - \rho(n))}},
\end{equation}
the same way can be applied to case $\mathtt{k=1}$ and $\mathtt{k=3}$. As $n$ increases, $\rho(n)$ typically decreases, before moderately increasing due to the cyclostationarity of solar irradiance. This reflects a weakening of temporal autocorrelation over short to medium horizons. Consequently, the denominator grows, making the $\mathtt{NICE^2}$ metric more sensitive to errors at longer horizons. However, for extended horizons, the assumption of stationarity becomes increasingly tenuous as non-stationary effects (\textit{e.g.}, seasonal trends, long-term variability) become dominant. In such cases, alternative normalization strategies or model recalibration could be necessary to ensure meaningful forecasting performance assessments.

\subsection{The \texttt{NICE}\ensuremath{^\Sigma} Metric: A Weighted Combination of Errors}
The $\mathtt{NICE^\Sigma}$ metric generalizes the $\mathtt{NICE^k}$ framework by combining multiple error norms into a single measure. It is defined as:
\begin{equation}
    \mathtt{NICE^\Sigma} = w_1  \times \mathtt{NICE^1} + w_2  \times \mathtt{NICE^2} + w_3 \times  \mathtt{NICE^3},
\end{equation}
where $w_1$, $w_2$, and $w_3$ are non-negative weights satisfying:
\begin{equation}
    w_1 + w_2 + w_3 = 1, \quad w_i \geq 0 \quad \forall i.
\end{equation}
Each component captures a different aspect of forecasting errors. The term $\mathtt{NICE^1}$, derived from the $\mathtt{L}^1$-norm, provides robustness to outliers by equally weighting all error magnitudes. The term $\mathtt{NICE^2}$, based on the $\mathtt{L}^2$-norm, emphasizes larger deviations while balancing general performance. Finally, $\mathtt{NICE^3}$, rooted in the $\mathtt{L}^3$-norm, amplifies rare, extreme forecasting errors, making it suitable for applications where large deviations carry significant risk. The flexibility of $\mathtt{NICE^\Sigma}$ lies in the choice of weights, allowing for evaluations adapted to specific application needs.
For instance, a balanced configuration might assign equal weights to all components:
\begin{equation}
    \mathtt{NICE^\Sigma} = \frac{1}{3}   \mathtt{NICE^1} + \frac{1}{3}   \mathtt{NICE^2} + \frac{1}{3}   \mathtt{NICE^3}.
\end{equation}
Alternatively, in risk-sensitive contexts, a higher weight can be assigned to $\mathtt{NICE^3}$ to better capture extreme deviations, as they may have critical implications in applications such as grid stability or financial risk assessment.
In conclusion, the $\mathtt{NICE^\Sigma}$ metric bridges the gap between different $\mathtt{NICE^n}$ metrics, offering a flexible tool to assess model performance across various forecasting scenarios. Proper calibration of weights ensures that $\mathtt{NICE^\Sigma}$ reflects both robustness to outliers and sensitivity to critical deviations, making it a versatile and interpretable metric for time series forecasting. In this study, an iso-weight definition is chosen to maintain neutrality, prevent bias toward a specific error type, and provide a balanced assessment of forecasting performance. An Interpretation and a physical sense of metrics related to $\mathtt{NICE^k}$ metrics are available in \ref{annex:interpretation}

%
%

\subsection{Model Evaluation Datasets}
To assess the performance of the proposed predictors, two distinct types of datasets are employed: synthetic time series and measurements from irradiance sensors. \cyr{Among the forecasting models evaluated in this study, only the $\mathtt{EL}$ (Extreme Learning) model involves a tunable hyperparameter: the regularization coefficient $\lambda$. This parameter appears in the computation of the output weights associated with a random projection of the input space onto a high-dimensional feature space. The regularization term $\lambda$ ensures numerical stability during the solution of the linear system involving these representations. In our experiments, The regularization coefficient $\lambda$ was selected by grid search on a logarithmic scale over the range $[10^{-6}, 10^{-1}]$, using in-sample $\mathtt{RMSE}$ as the selection criterion. Optimal values were typically close to $\lambda = 10^{-3}$, and sensitivity tests confirmed that predictive performance remained stable over this range. The remaining models ($\mathtt{AR}, \mathtt{SP}$, etc.) do not require hyperparameter tuning.}

\subsubsection{Synthetic Data}
\cyr{Both theoretical and empirical $\mathtt{NICE}$ metrics are evaluated for three forecasting models \cite{VOYANT2022747}, based on 100 Monte Carlo simulations using synthetic time series of 50,000 data points each}:  
(i) $\mathtt{P}$ (Persistence), which assumes that the future value equals the current one, serving as a naive baseline;  
(ii) $\mathtt{SP}$ (Smart Persistence), which extends the basic persistence approach by incorporating daily or periodic adjustments (\textit{e.g.}, diurnal normalization);  
(iii) $\mathtt{AR}$ (AutoRegressive), a linear model that predicts future values based on a weighted combination of past observations.
\cyr{A wide range of periodic and synthetic signals is based on cyclostationary processes combining a deterministic clear-sky baseline (modeled as a non-negative sine wave) and additive Gaussian noise. Temporal smoothness is controlled by a moving average filter of randomized length, and key parameters—such as the period, time step, noise standard deviation, and filter kernel size—are independently sampled from predefined distributions (see \cite {VOYANT2026123913} for information). This synthetically-induced variability allows a systematic assessment of the sensitivity and robustness of both theoretical and empirical $\mathtt{NICE^k}$ metrics under diverse solar-like dynamics}.

\subsubsection{Measured Data}
\begin{figure}
    \centering
    \hspace{-0.5cm}
    \begin{subfigure}[b]{0.45\textwidth} 
        \centering
        \includegraphics[width=\textwidth]{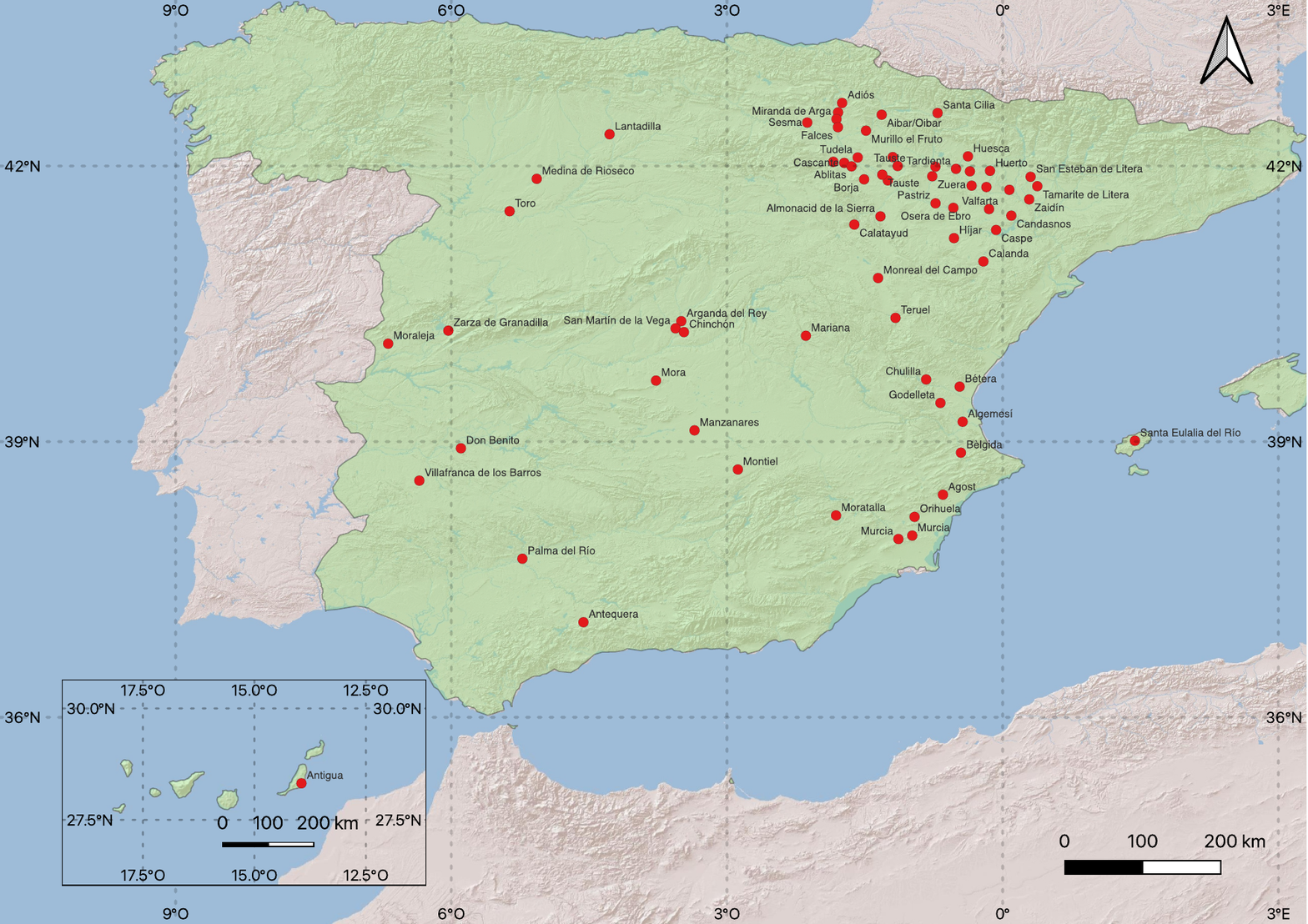}
        \captionsetup{justification=centerlast, singlelinecheck=false}
        \caption{Sensors Locations}
        \label{fig:locations}
    \end{subfigure}
    \hfill
    \begin{subfigure}[b]{0.54\textwidth} 
        \centering
        \raisebox{0.5cm}{ 
        \includegraphics[width=\textwidth]{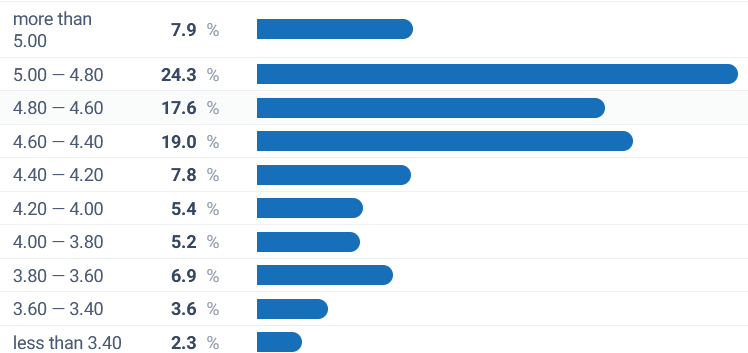}}
        \caption{$\mathtt{I}$ Distribution in Spain (kWh/m²)}
        \label{fig:distribution}
    \end{subfigure}
    \caption{\texttt{SIAR} DATA as Benchmark. 
    Statistical summary: Average: 4.49 kWh/m², 
    Max: 5.03 kWh/m², Min: 3.55 kWh/m², 
    Median (P50): 4.58 kWh/m², 
    Percentile 90: 4.95 kWh/m², 
    Percentile 75: 4.85 kWh/m², 
    Percentile 25: 4.23 kWh/m², 
    Percentile 10: 3.72 kWh/m².}
    \label{fig:SIAR_DATA}
\end{figure}

The second set of simulations and evaluations was carried out utilizing experimental data obtained from the Agroclimatic Information System for Irrigation (\texttt{SIAR}) network, which is managed by the Spanish Ministry of Agriculture, Fisheries, Food, and the Environment. The dataset comprises measurements of Global Horizontal Irradiance ($\mathtt{I}$) obtained from 68 meteorological stations located throughout mainland Spain and the Canary Islands. The spatial distribution of these meteorological stations is shown in Figure~\ref{fig:SIAR_DATA}. Detailed information regarding these locations can be found in reference \cite{despotovic2024}. In addition to the previous models, two complementary approaches are considered:  
(i) $\mathtt{ARTU}$, an $\mathtt{AR(2)\text{-}like}$ model without a learning phase that derives its name from the pronunciation of AR-two \cite{VOYANT2022747}; and (ii) $\mathtt{EL}$ (Extreme Learning), which extends traditional neural-based methods through a fast and efficient training procedure \cite{despotovic2024,VOYANT2025113490}.
\cyr{No post-processing or bias correction was applied to model outputs. Forecasts were evaluated in their raw form to ensure a neutral comparison based solely on predictive skill.}

Geographically, the sites are spread out over a large range of latitude, longitude, and altitude, incorporating variability characteristic of Spanish distinctive climate. This comprehensive and geographically varied dataset facilitates thorough modeling and examination of solar irradiance patterns, integrating multiple climatic and topographical factors throughout Spain. 
The measurements span from January 1, 2017, to December 31, 2020, at a 30-minute resolution, providing a comprehensive dataset that captures seasonal variations. \cyr{If no pre-processing is required for synthetic data, as the time series are generated from known stochastic processes with controlled statistical properties, the situation differs for the \texttt{SIAR} irradiance measurements, which underwent rigorous quality control combining automated filtering and expert visual inspection procedures, as described in \cite{forstinger:hal-03657585}. This process ensures that only physically consistent and high-confidence data are retained. Forecasting models were applied directly to raw irradiance signals, without normalization based on clear-sky models. This approach follows the "clearsky-free" paradigm introduced in \cite{VOYANT2025113490}, which avoids injecting uncertainty from auxiliary models and preserves the native temporal variability \cite{VOYANT2026123913}, a critical aspect for time-series learning models such as $\mathtt{EL}$.
}

%
%

\section{Results}
\label{sec:results}
Results are provided for two types of simulations. The first involves a controlled environment with synthetic time series whose characteristics are fully defined, while the second involves experimental stations with $\mathtt{I}$ measurements, accounting for all uncertainties related to the clear sky model, measurements, and their timestamps. \cyr{All results reported in this section are based on a strict 75/25 split between in-sample (training) and out-of-sample (testing) data. For each forecasting model, the first half of the time series was used for calibration, and the second half was reserved for independent evaluation. This protocol was uniformly applied to both synthetic and empirical datasets.}

%
%

\subsection{Monte Carlo-Based Analysis of Forecasting Metrics}
To rigorously evaluate the properties of the proposed metrics under controlled conditions, a Monte Carlo framework was implemented. This approach enables the simulation of synthetic forecast–observation pairs under various statistical assumptions, allowing both theoretical and empirical metrics to be compared systematically. The objective is twofold: first, to assess whether the theoretical formulations hold under ideal conditions, and second, to evaluate the discriminative power of each metric when applied to competing forecasting models.

\subsubsection{Validation of the Theory as a Consequence of Statistical Assumptions}
\begin{figure}[h!]
    \centering
    \includegraphics[width=0.75\textwidth]{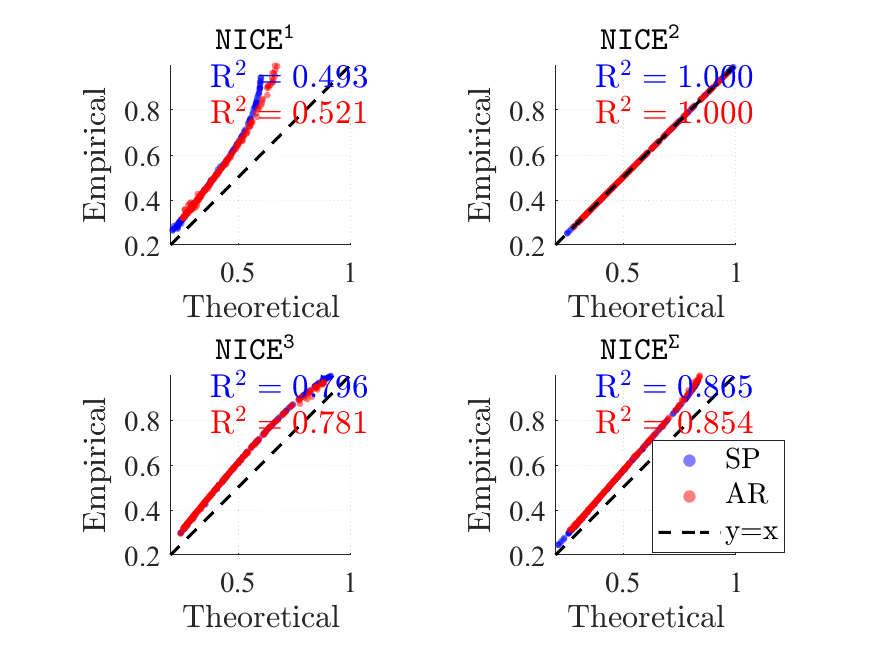}
    \caption{Comparison of Theoretical and Empirical $\mathtt{NICE}$ Metrics for Different Models. The perfect match observed in $\mathtt{NICE^2}$ (top-right) validates the theoretical assumptions for variance, while the other metrics exhibit discrepancies due to the limitations of the normality assumption. Both models tested ($\mathtt{SP}$ and $\mathtt{AR}$) showed no discontinuities during the simulations, highlighting their stability.}
    \label{fig:nice_metrics}
\end{figure}

The goal was to test the validity of key assumptions, namely stationarity (the stability of statistical properties over time) and normality (a simplifying assumption often used in theoretical derivations). 
The results, illustrated in Figure~\ref{fig:nice_metrics}, demonstrate a perfect match between theoretical and empirical metrics for $\mathtt{NICE^2}$ ($\mathtt{R}^2 = 1.0$), confirming that the second-order moment (variance) is well modeled under the theoretical assumptions. The assumption of local stationarity at lag $n = 1$  used for the theoretical derivation of $\mathtt{NICE^2}$ is validated by this figure. The first-order weak stationarity hypothesis assumes that the expectation $\mathbb{E}[y(t)]$ remains constant for all $t$, but only over a short horizon (between $t$ and $t+\Delta t$). Given this limited time scale, this assumption is reasonable and generally holds in practical applications.
However, for $\mathtt{NICE^1}$, $\mathtt{NICE^3}$, and $\mathtt{NICE^\Sigma}$, discrepancies were observed. These deviations suggest that the normality assumption does not hold well in these cases, and empirical metrics provide more robust and realistic estimates.
It is also worth noting that no discontinuities were observed across the Monte Carlo simulations for both models tested ($\mathtt{SP}$ and $\mathtt{AR}$). This highlights the stability of the predictors in the presence of varying input condifurther supporting the validity of the empirical approach, 
Monte Carlo simulations were essential for validating theoretical assumptions, identifying the limitations of theoretical models, and demonstrating the robustness and simplicity of empirical metrics, which are more suitable for complex datasets.
Based on these findings, future analyses will exclusively rely on empirical approaches, as they are simpler to compute, robust to restrictive assumptions, and better aligned with  observations.

\subsubsection{Benchmarks of Errors Metrics with Synthetic Data}
\begin{figure}
    \centering
    \includegraphics[width=\textwidth]{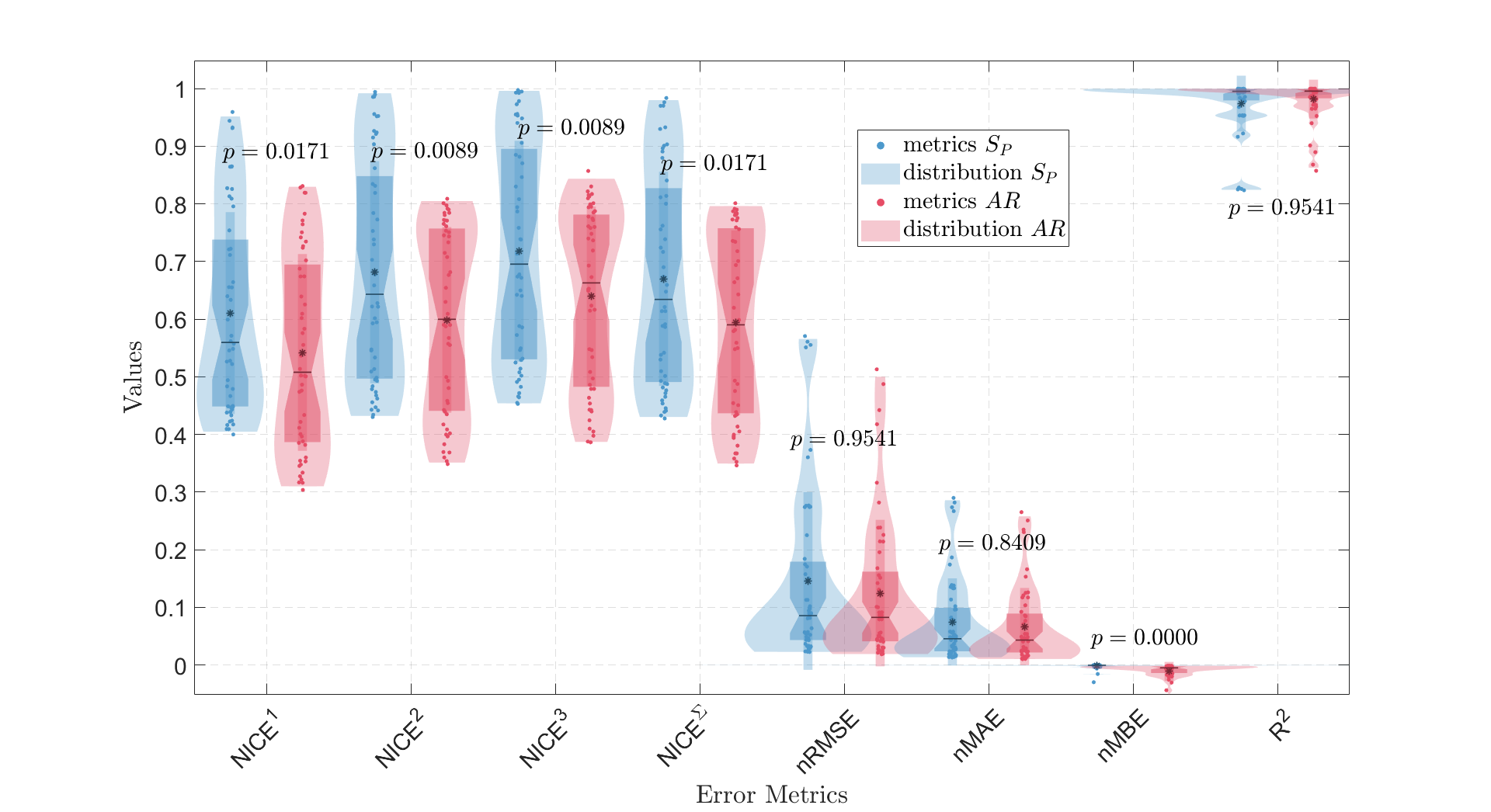}
    \caption{Distribution of normalized error metrics for $\mathtt{SP}$ (smart persistence) and $\mathtt{AR}$ (autoregressive) models across 100 simulations. The $p$-values from the \texttt{Kolmogorov-Smirnov} test, displayed above the violins, provide a statistical comparison between the distributions for $\mathtt{SP}$ and $\mathtt{AR}$. Low $p$-values indicate significant differences, while higher $p$-values suggest similarity.}
    \label{fig:violin_metrics}
\end{figure}

The violin plot (Figure~\ref{fig:violin_metrics}) illustrates the distribution of normalized error metrics for two predictors: $\mathtt{SP}$ (smart persistence) and $\mathtt{AR}$. The metrics include $\mathtt{NICE^1}$, $\mathtt{NICE^2}$, $\mathtt{NICE^3}$, $\mathtt{NICE^\Sigma}$, $\mathtt{nRMSE}$, $\mathtt{nMAE}$, $\mathtt{nMBE}$, and $\mathtt{R^2}$, derived from simulations designed to evaluate the comparative performance of these predictors. Statistical comparisons between the distributions were performed using the \texttt{Kolmogorov-Smirnov} test, with $p$-values displayed above each pair of violins.
The results reveal that $\mathtt{R^2}$, $\mathtt{nMAE}$, and $\mathtt{nRMSE}$ fail to differentiate between the two predictors, as their $p$-values exceed the 0.05 significance threshold. This suggests that these conventional metrics are inadequate for identifying nuanced differences between $\mathtt{SP}$ and $\mathtt{AR}$. Conversely, the $\mathtt{NICE}$ family of metrics ($\mathtt{NICE^1}$, $\mathtt{NICE^2}$, $\mathtt{NICE^3}$, and $\mathtt{NICE^\Sigma}$) consistently distinguish the predictors, with $p < 0.05$ across all cases. Among these, $\mathtt{NICE^\Sigma}$ demonstrates enhanced robustness by integrating multiple error components, offering a more comprehensive and extended perspective. 
Notably, $\mathtt{NICE}$ metrics provide a clearer interpretation of relative predictor performance. For instance, $\mathtt{SP}$ achieves approximately 65\% of the maximal error, while $\mathtt{AR}$ reaches 60\%, resulting in a five-percentage-point difference that is not evident with conventional metrics. This highlights the utility of $\mathtt{NICE}$ metrics in revealing subtle yet impactful variations in predictive accuracy.
Although $\mathtt{nMBE}$ shows a significant difference between the predictors ($p < 0.05$), its magnitude remains minimal (less than 1\%), indicating that the bias is negligible for both models. Consequently, $\mathtt{nMBE}$ serves more as a necessary condition for predictor reliability than as a discriminatory metric. 
Interestingly, the distribution shapes further elucidate the behavior of these metrics. Unlike conventional metrics, which exhibit Gaussian-like distributions, $\mathtt{NICE}$ metrics display nearly uniform distributions, reflecting a broader dispersion of errors. This suggests a reduction in degeneracy, allowing $\mathtt{NICE}$ metrics to better capture variations in prediction quality across simulations. 
Overall, these findings emphasize the superiority of $\mathtt{NICE}$ metrics for robust predictor comparison and underscore their interpretability and sensitivity to differences that conventional metrics fail to detect.

%
%

\subsection{SIAR-Based Forecast Metric Analysis}
To complement the Monte Carlo-based results, the analysis was applied to the \texttt{SIAR} dataset, which offers detailed irradiance measurements under diverse sky conditions. This dataset enables a structured comparison of forecasting models using both classical and $\mathtt{NICE}$ metrics. The objective is to evaluate the consistency, discriminative power, and statistical robustness of the metrics when applied to measured time series, and to verify whether previous observations hold across different models and horizons.

\subsubsection{Discriminative Power of $\texttt{NICE}$ Metrics in Forecasting Model Evaluation}
\begin{figure}[h!]
    \centering
    \includegraphics[width=0.9\textwidth]{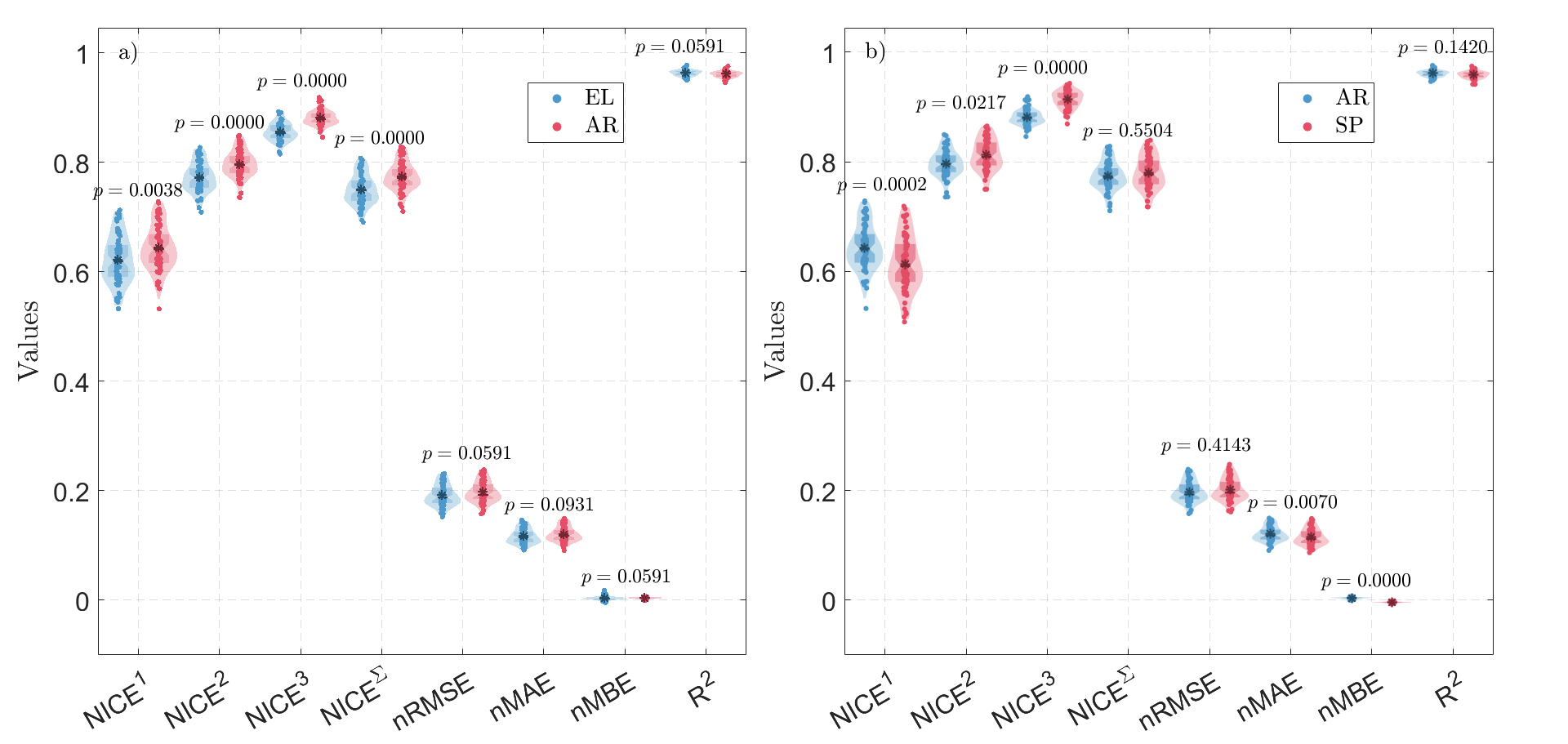}
    \caption{Distribution of error metrics $\mathtt{NICE^1}$, $\mathtt{NICE^2}$, $\mathtt{NICE^3}$, $\mathtt{NICE^\Sigma}$, $\mathtt{nRMSE}$, $\mathtt{nMAE}$, $\mathtt{nMBE}$, and $\mathtt{R^2}$ for the \texttt{SIAR} dataset. (a) Comparison of metrics for the $\mathtt{EL}$ (Extreme Learning Machine) and $\mathtt{AR}$ (autoregressive) models. (b) Comparison of the same metrics for the $\mathtt{AR}$ (autoregressive) and $\mathtt{SP}$ (smart persistence) models. Statistical significance between the models is indicated by \textit{p-values} from the \texttt{Kolmogorov-Smirnov} test for each pair of violins.}
    \label{fig:SIAR_violin}
\end{figure}
\begin{figure}
    \centering
    \includegraphics[width=0.9\textwidth]{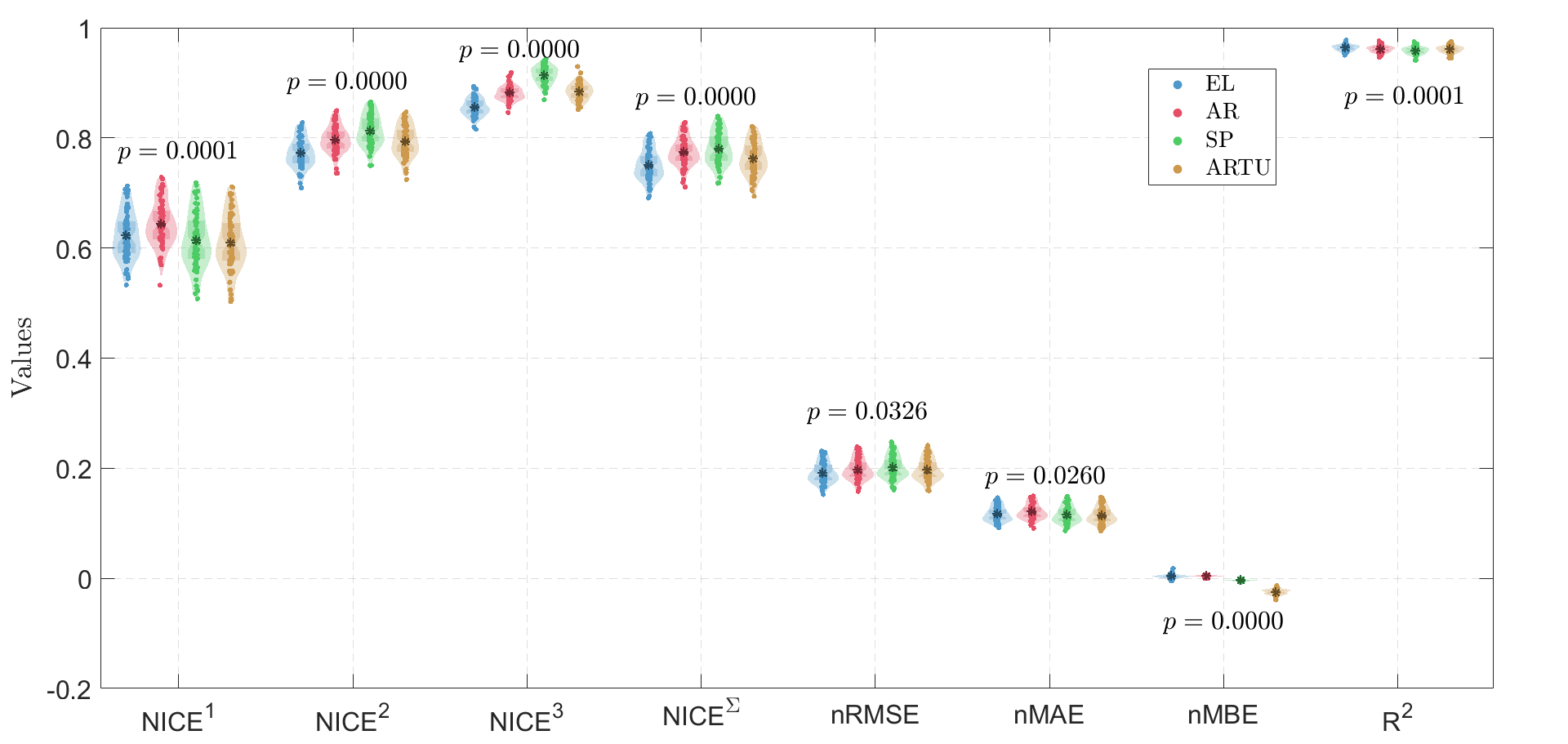}
    \caption{Comparison of error metrics $\mathtt{NICE^1}$, $\mathtt{NICE^2}$, $\mathtt{NICE^3}$, $\mathtt{NICE^\Sigma}$, $\mathtt{nRMSE}$, $\mathtt{nMAE}$, $\mathtt{nMBE}$, and $\mathtt{R^2}$ for \texttt{SIAR} dataset for four forecasting models: $\mathtt{EL}$ (Extreme Learning Machine), $\mathtt{AR}$ (Autoregressive), $\mathtt{SP}$ (Smart Persistence), and $\mathtt{ARTU}$ (Particular Autoregressive Model of Order Two \cite{VOYANT2022747}). Statistical significance between the models was assessed using the Kruskal-Wallis test, with \textit{\textit{p-values}} provided for each group of violins.}
    \label{fig:SIAR_violin_FourModels}
\end{figure}

The results shown in Figure~\ref{fig:SIAR_violin} provide a comparative analysis of error metrics for solar irradiance dataset, calculated for forecasting horizon of 30 min. The violin plots display the distribution of error metrics $\mathtt{NICE^1}$, $\mathtt{NICE^2}$, $\mathtt{NICE^3}$, $\mathtt{NICE^\Sigma}$, $\mathtt{nRMSE}$, $\mathtt{nMAE}$, $\mathtt{nMBE}$, and $\mathtt{R^2}$, for three different models: $\mathtt{EL}$ (Extreme Learning Machine; \cite{despotovic2024}), $\mathtt{AR}$ (autoregressive), $\mathtt{SP}$ (smart persistence) and $\mathtt{ARTU}$ (particular autoregressive mode; \cite{VOYANT2022747}). 
It is apparent that the $\mathtt{NICE}$ family of metrics offers a more sensitive and robust assessment of model performance in comparison to conventional error metrics. The \textit{p-values}, derived from the \texttt{Kolmogorov-Smirnov} test, presented for each pair of violins, serve to indicate the presence of statistically significant differences between the models under comparison.
These values show that the $\mathtt{NICE}$ metrics are able to better capture subtle differences in forecasting accuracy, which is especially important when dealing with complex datasets such as solar irradiance.
$\mathtt{nRMSE}$, $\mathtt{nMAE}$, $\mathtt{nMBE}$, and $\mathtt{R^2}$ fail to differentiate between the predictors in both cases (\textit{e.g.}, $\mathtt{EL}$ vs.
$\mathtt{AR}$, $\mathtt{AR}$ vs. $\mathtt{SP}$), as their \textit{p-values} exceed the 0.05 significance threshold. Moreover, unlike conventional metrics ($\mathtt{nRMSE}$,
$\mathtt{nMAE})$, $\mathtt{NICE}$ family distributions are generally normal in the \texttt{Jarque-Bera} sense \cite{jarque1987test}, and notably more dispersed, which facilitates inter-study
comparisons by avoiding the overly similar error values that hinder generalizability. Similar conclusions arise from Figure~\ref{fig:SIAR_violin_FourModels}, which compares four forecasting models, including the newly introduced $\mathtt{ARTU}$ method (a specific $\mathtt{AR}$(2) model \cite{VOYANT2022747}). Model performance was assessed using cross-metric comparisons, with statistical significance tested via the \texttt{Kruskal-Wallis} method wich is a non-parametric alternative to one-way \texttt{ANOVA}, used to compare more than two independent samples when the assumptions of normality or homogeneity of variances are violated. Unlike the \texttt{Kolmogorov-Smirnov} test, which is
limited to comparing two distributions, \texttt{Kruskal-Wallis} allows for multi-group comparisons. For $\mathtt{nRMSE}$ and $\mathtt{nMAE}$, \textit{p-values} of 0.0326 and 0.0260 suggest limited discriminative power. In contrast, $\mathtt{NICE}$ metrics exhibit far stronger significance, with \textit{p} $< 0.001$, highlighting their superior sensitivity in detecting subtle differences across models.

\subsubsection{Horizon-Wise Discriminative Power of $\texttt{NICE}$ Versus Conventional Metrics}
The analysis of the same set of metrics for two forecasting models, $\mathtt{EL}$ and $\mathtt{AR}$, across multiple forecasting horizons (from 30 minutes to 6 hours, in 30-minute steps) has been conducted. Statistical significance between the models was assessed using the \texttt{Kolmogorov–Smirnov} test. For enhanced readability and to support quantitative interpretation, these results are summarized in Table~\ref{tab:medians_pvalues_horizons}, which includes both the \textit{p-values} and the medians of each metric for both models.
To streamline the analysis and avoid redundancy, we retain only one representative metric from the $\mathtt{NICE}$ family, namely $\mathtt{NICE^\Sigma}$, which has previously demonstrated a good sensitivity and robustness. Furthermore, $\mathtt{nMBE}$ and $\mathtt{R^2}$ are omitted from this comparison due to their limited impact on model discrimination.
\begin{table}[ht]
\centering
\begin{footnotesize}
\begin{tabular}{c l l l}
\toprule
\multicolumn{1}{|c|}{\textbf{$n\Delta t$}} & \multicolumn{1}{c|}{$\mathtt{NICE}^\Sigma$} & \multicolumn{1}{c|}{$\mathtt{nRMSE}$} & \multicolumn{1}{c|}{$\mathtt{nMAE}$} \\
\midrule
30 min & $\mathtt{EL}$: 0.75, $\mathtt{AR}$: 0.77 \; \textbf{(<0.001)} & $\mathtt{EL}$: 0.19, $\mathtt{AR}$: 0.19 \; (0.0591) & $\mathtt{EL}$: 0.11, $\mathtt{AR}$: 0.12 \; (0.0931) \\
1.0 h  & $\mathtt{EL}$: 0.61, $\mathtt{AR}$: 0.64 \; \textbf{(<0.001)} & $\mathtt{EL}$: 0.24, $\mathtt{AR}$: 0.25 \; \textbf{(0.0070)} & $\mathtt{EL}$: 0.15, $\mathtt{AR}$: 0.16 \; \textbf{(0.0126)} \\
1.5 h  & $\mathtt{EL}$: 0.51, $\mathtt{AR}$: 0.54 \; \textbf{(<0.001)} & $\mathtt{EL}$: 0.26, $\mathtt{AR}$: 0.28 \; \textbf{(<0.001)} & $\mathtt{EL}$: 0.17, $\mathtt{AR}$: 0.19 \; \textbf{(<0.001)} \\
2.0 h  & $\mathtt{EL}$: 0.45, $\mathtt{AR}$: 0.48 \; \textbf{(<0.001)} & $\mathtt{EL}$: 0.28, $\mathtt{AR}$: 0.30 \; \textbf{(<0.001)} & $\mathtt{EL}$: 0.19, $\mathtt{AR}$: 0.21 \; \textbf{(<0.001)} \\
2.5 h  & $\mathtt{EL}$: 0.41, $\mathtt{AR}$: 0.43 \; \textbf{(<0.001)} & $\mathtt{EL}$: 0.30, $\mathtt{AR}$: 0.32 \; \textbf{(<0.001)} & $\mathtt{EL}$: 0.20, $\mathtt{AR}$: 0.22 \; \textbf{(<0.001)} \\
3.0 h  & $\mathtt{EL}$: 0.37, $\mathtt{AR}$: 0.40 \; \textbf{(<0.001)} & $\mathtt{EL}$: 0.31, $\mathtt{AR}$: 0.33 \; \textbf{(<0.001)} & $\mathtt{EL}$: 0.21, $\mathtt{AR}$: 0.23 \; \textbf{(0.0010)} \\
3.5 h  & $\mathtt{EL}$: 0.35, $\mathtt{AR}$: 0.37 \; \textbf{(<0.001)} & $\mathtt{EL}$: 0.32, $\mathtt{AR}$: 0.34 \; \textbf{(<0.001)} & $\mathtt{EL}$: 0.22, $\mathtt{AR}$: 0.23 \; \textbf{(0.0126)} \\
4.0 h  & $\mathtt{EL}$: 0.33, $\mathtt{AR}$: 0.35 \; \textbf{(<0.001)} & $\mathtt{EL}$: 0.33, $\mathtt{AR}$: 0.35 \; \textbf{(<0.001)} & $\mathtt{EL}$: 0.23, $\mathtt{AR}$: 0.24 \; \textbf{(0.0364)} \\
4.5 h  & $\mathtt{EL}$: 0.32, $\mathtt{AR}$: 0.33 \; \textbf{(<0.001)} & $\mathtt{EL}$: 0.33, $\mathtt{AR}$: 0.36 \; \textbf{(<0.001)} & $\mathtt{EL}$: 0.23, $\mathtt{AR}$: 0.24 \; (0.1420) \\
5.0 h  & $\mathtt{EL}$: 0.31, $\mathtt{AR}$: 0.32 \; \textbf{(0.0020)} & $\mathtt{EL}$: 0.34, $\mathtt{AR}$: 0.36 \; \textbf{(0.0020)} & $\mathtt{EL}$: 0.24, $\mathtt{AR}$: 0.24 \; (0.3000) \\
5.5 h  & $\mathtt{EL}$: 0.30, $\mathtt{AR}$: 0.31 \; \textbf{(0.0070)} & $\mathtt{EL}$: 0.34, $\mathtt{AR}$: 0.36 \; \textbf{(0.0020)} & $\mathtt{EL}$: 0.24, $\mathtt{AR}$: 0.25 \; (0.6984) \\
6.0 h  & $\mathtt{EL}$: 0.29, $\mathtt{AR}$: 0.31 \; \textbf{(0.0038)} & $\mathtt{EL}$: 0.35, $\mathtt{AR}$: 0.37 \; \textbf{(0.0020)} & $\mathtt{EL}$: 0.24, $\mathtt{AR}$: 0.25 \; (0.6984) \\
\bottomrule
\end{tabular}
\end{footnotesize}
\caption{Medians for $\mathtt{EL}$ and $\mathtt{AR}$ models and associated $p$-values for selected error metrics over forecasting horizons. Bold cases indicate statistical significance ($p < 0.05$).}
\label{tab:medians_pvalues_horizons}
\end{table}

This table presents the evolution of \textit{p-values} and medians for $\mathtt{NICE^\Sigma}$, $\mathtt{nRMSE}$, and $\mathtt{nMAE}$ over increasing forecast horizons ($n\Delta t$ from 30 minutes to 6 hours). While $\mathtt{nRMSE}$ becomes statistically significant only from the 1-hour horizon onward ($p = 0.0070$), $\mathtt{nMAE}$ fails to reach significance at both 30 minutes and 6 hours ($p = 0.0931$ and $p = 0.6984$ respectively), highlighting its limited reliability for long-term comparison.
In contrast, $\mathtt{NICE^\Sigma}$ consistently shows strong statistical significance across all horizons and remains below the 0.05 threshold even at 6 hours. Its medians decrease gradually from 0.75 to 0.29 (for $\mathtt{EL}$) and from 0.77 to 0.31 (for $\mathtt{AR}$), clearly reflecting the degradation of model accuracy with time. On the other hand, $\mathtt{nRMSE}$ and $\mathtt{nMAE}$ exhibit relatively stable median values (\textit{e.g.}, $\mathtt{nRMSE}$: from 0.19 to 0.37), which limits their discriminative resolution.
These results underline that $\mathtt{NICE^\Sigma}$ not only captures subtle differences more effectively but also provides a horizon-independent and interpretable framework for model evaluation, simplifying the analysis without compromising statistical robustness. \cyr{These comparative results echo prior findings in the literature. In particular, \cite{zhang2015suite} emphasized the sensitivity of solar forecast metrics to normalization procedures and evaluation protocols, while \cite{foley2012current} reported similar limitations of single-metric evaluations in wind forecasting. The proposed $\mathtt{NICE^k}$ metrics address these issues by enabling structured, norm-based comparisons that remain stable across datasets and forecasting horizons.}

\subsection{Practical Guidance for \texttt{NICE} Metric Selection}
The results presented above highlight the relevance of using the $\mathtt{NICE}^\Sigma$ composite metric as a practical guidance tool in multidimensional forecast evaluation. Each individual metric ($\mathtt{NICE}^1$, $\mathtt{NICE}^2$, and $\mathtt{NICE}^3$) captures a distinct aspect of the forecasting error: trend fidelity, penalization of large deviations, and sensitivity to outliers, respectively. While it may be difficult to recommend a single metric as universally optimal, the blended $\mathtt{NICE}^\Sigma$ offers a balanced representation across these dimensions. Although equal weights were used in this study for simplicity and reproducibility, they can be adapted according to the operational context. For instance, isotropic weighting may suit general irradiance estimation tasks, while territorial forecasting could benefit from emphasizing $\mathtt{NICE}^1$ and $\mathtt{NICE}^2$. In contrast, applications such as smart grid management where the economic impact of extreme errors is significant, might justify a higher weight for $\mathtt{NICE}^3$. Furthermore, the observed lack of discrimination between methods for shorter horizons (\textit{e.g.}, $\mathtt{AR}$ vs $\mathtt{SP}$ under $\mathtt{NICE}^\Sigma$) should not be seen as a limitation of the metric, but rather as a reflection of the inherent loss of predictability in chaotic systems beyond the \texttt{Lyapunov} time scale \cite{PhysRevX.13.021018}. This convergence in performance is expected and less impactful in practice, especially when short-horizon accuracy remains the key for most energy applications. Hence, $\mathtt{NICE}$ metrics can serve both as evaluation tools and as operational decision aids, provided they are interpreted within the appropriate temporal and contextual framework.

%
%

\section{Conclusion}
\label{sec:conclusion}
The objective of this study was to address limitation in evaluating solar irradiance forecasting models by proposing and validating a new family of error metrics called \cyr{$\mathtt{NICE^k}$} (Normalized Informed Comparison of Errors) and comparing their relative merit to conventional error metrics.The scheme integrates error information across multiple norm spaces ($\mathtt{L}^1$, $\mathtt{L}^2$, and $\mathtt{L}^3$), while normalizing with respect to intrinsic variability and temporal predictability of the data. The study used synthetic Monte Carlo trials and data from Spain \texttt{SIAR} network, which included 68 meteorological stations in various climatic and geographic environments. The synthetic simulations tested theoretical hypotheses (normality and stationarity) and benchmarked predictors, while the measured data analysis evaluated more advanced techniques. The Monte Carlo experiments proved theoretical \cyr{$\mathtt{NICE^k}$} measures to perfectly match only under rigorous statistical assumptions (\textit{e.g.}, normality), as in $\mathtt{NICE^2}$ ($\mathtt{R}^2 = 1.0$). The discrepancies found in $\mathtt{NICE^1}$, $\mathtt{NICE^3}$, and $\mathtt{NICE^\Sigma}$ highlighted theoretical derivation limitations, supporting the superiority of empirical measures. Interestingly, \cyr{$\mathtt{NICE^k}$} metrics outcompeted standard measures in differentiating among predictors at all investigated data points, and $\mathtt{NICE^\Sigma}$ proved exceptionally robust by combining trend fidelity and outlier sensitivity. Conventional metrics such as $\mathtt{nRMSE}$ and $\mathtt{nMAE}$ tended to fail in detecting meaningful differences ($p > 0.05$), highlighting their reduced discriminative power.
Analysis of \texttt{SIAR} data reinforced these conclusions. 
\cyr{$\mathtt{NICE}$ measures consistently distinguished models (\textit{e.g.}, Extreme Learning ($\mathtt{EL}$) vs. Autoregressive models ($\mathtt{AR}), \mathtt{AR}$ vs. Smart Persistence ($\mathtt{SP}$)
) for different forecasting horizons (30 minutes to 6 hours), displaying statistical significance ($p < 0.05$) even when conventional measures failed. These models were deliberately selected to cover a spectrum of error dynamics (from linear stochastic behavior to machine learning-based approximations) thus demonstrating the ability of $\mathtt{NICE^k}$ to adapt to various forecast architectures and error profiles.} $\mathtt{NICE^\Sigma}$ illustrated consistent sensitivity as median values decreased gradually from 0.75 to 0.29 ($\mathtt{EL}$) and 0.77 to 0.31 ($\mathtt{AR}$) at extended horizons, indicating systematic degradation in accuracy. In contrast, $\mathtt{nRMSE}$ and $\mathtt{nMAE}$ demonstrated stable medians (\textit{e.g.}, 0.19--0.37 for $\mathtt{nRMSE}$), suppressing essential variations in performance. \texttt{Kruskal-Wallis} tests reinforced superiority of $\mathtt{NICE}$ measures with $p$-values $< 0.001$ in contrast to poorer significance for $\mathtt{nRMSE}$ ($p = 0.0326$) and $\mathtt{nMAE}$ ($p = 0.0260$).
\cyr{This research establishes the $\mathtt{NICE^k}$ family (particularly $\mathtt{NICE^\Sigma}$) as a theoretically grounded and practically robust framework for evaluating deterministic solar forecasts, capable of outperforming conventional metrics in both interpretability and discriminative power.} \cyr{In general they are better than conventional metrics due to their resilience to limiting hypotheses}, sensitivity to refined differences in models, and flexibility to operational environments. Conventional measures can still be useful for simple comparisons, but $\mathtt{NICE}$ measures provide an overall system for sophisticated uses. \cyr{For instance, under certain horizons (almost all), $\mathtt{nRMSE}$ yields nearly identical scores for $\mathtt{AR}$ and $\mathtt{EL}$ models, despite their differing error profiles, a contrast effectively highlighted by $\mathtt{NICE^k}$.} Overall, the goal is to establish $\mathtt{NICE^k}$ as a standard, adaptive framework for evaluating solar forecasting accuracy, thereby supporting innovation in the field and contributing to the global transition toward more sustainable energy systems.
\cyr{Beyond pure model evaluation, the $\mathtt{NICE^k}$ framework can support operational energy forecasting tasks, including performance benchmarking under varying cloud regimes, sensitivity analyses in grid simulations, or reliability scoring in hybrid forecasting chains.} \cyr{By enabling more accurate and discriminative forecast evaluation, the $\mathtt{NICE^k}$ framework can support improved dispatch planning, optimized battery scheduling in hybrid PV-storage systems, and enhanced reserve allocation, thereby increasing grid stability and reducing operational costs.} \cyr{From an implementation perspective, $\mathtt{NICE^k}$ metrics do not introduce additional computational burden compared to conventional metrics, as they rely solely on standard vector norms and a data-driven normalization based on temporal variability. Their design ensures that improved discrimination does not come at the expense of usability or efficiency.}

Future research should focus on expanding the empirical validation of $\mathtt{NICE^k}$ across a wider range of forecasting models and operational conditions, and on integrating these metrics into energy decision-making workflows.

\section*{Acknowledgements}
This research was partially funded by the $\mathtt{ANR}$ under grant $\mathtt{SAPHIR}$ project ($\texttt{ANR}$ reference: $\mathtt{Sensor Augmented weather Prediction at HIgh-Resolution [ANR-21-CE04-0014}]$), whose support is gratefully acknowledged. Authors thank $\mathtt{Elsevier}$ and the $\mathtt{Scopus}$ platform for providing structured access to bibliographic data. Their $\mathtt{API}$ and metadata services were essential to the successful implementation of the text mining pipeline presented in this study.

%
%
 
\appendix
\section{Properties of \texttt{NICE}$^{k}$ Error Metrics Family}
\label{annex:properties}
The $\mathtt{NICE^2}$ metric is invariant to scale transformations. To demonstrate this, consider a scaled version of the time series and its predictions: $y(t)' = \eta \,  y(t)$ and $\hat{y}(t)' = \eta   \hat{y}(t)$, where $\eta > 0$. Let $\mathtt{RMSE_X'}$ denote the $\mathtt{RMSE}$ of model $\mathtt{X}$ on the scaled data, and $\mathtt{RMSE_P'}$ denote the $\mathtt{RMSE}$ of the persistence model on the scaled data. Then:
\begin{align*}
    \mathtt{RMSE_X'} &= \sqrt{\frac{1}{n} \sum_{t=1}^n (y(t)' - \hat{y}(t)')^2} = \sqrt{\frac{1}{n} \sum_{t=1}^n \eta^2 (y(t) - \hat{y}(t))^2} = \eta   \mathtt{RMSE_X}.
\end{align*}
Similarly, for the persistence model, $\mathtt{RMSE_P'} = \eta \,  \mathtt{RMSE_P}$.
Therefore, the $\mathtt{NICE^2}$ for the scaled data is:
\begin{equation*}
    \frac{\mathtt{RMSE_X'}}{\mathtt{RMSE_P'}} = \frac{\eta \,  \mathtt{RMSE_X}}{\eta \,  \mathtt{RMSE_P}} = \frac{\mathtt{RMSE_X}}{\mathtt{RMSE_P}} = \mathtt{NICE^2}.
\end{equation*}
This demonstrates that $\mathtt{NICE^2}$ is invariant to scale transformations. Another important property is its invariance to translations, which ensures that shifting all values by a constant does not affect the metric.
If $y(t)' = y(t) + c$ and $\hat{y}(t)' = \hat{y}(t) + c$ for some constant $c$, then:
\begin{align*}
    (y(t)' - \hat{y}(t)') &= (y(t) + c) - (\hat{y}(t) + c) = y(t) - \hat{y}(t), \\
    (y(t)' - y(t-\Delta t)') &= (y(t) + c) - (y(t-\Delta t) + c) = y(t) - y(t-\Delta t).
\end{align*}
Since both $(y(t) - \hat{y}(t))$ and $(y(t) - y(t-\Delta t))$ remain unchanged under translation, neither $\mathtt{RMSE_X}$ nor $\mathtt{RMSE_P}$ is affected. Consequently, $\mathtt{NICE^2}$ remains invariant to translations, reinforcing its robustness as a scale-free and shift-independent evaluation metric. After some mathematical manipulations, same conclusions are obtained for other cases, allowing to validate properties for $\mathtt{NICE^k}$ \ ($\forall k$).

\section{Derivation of \texttt{NICE\ensuremath{^1}} and \texttt{NICE\ensuremath{^3}}}
\label{NICE1&2}
The normalized metrics $\mathtt{NICE^1}$ and $\mathtt{NICE^3}$ are derived to evaluate forecasting accuracy relative to a persistence model. Both metrics depend on the variance $\sigma^2$ of the time series and the lag-1 autocorrelation coefficient $\rho(1)$. Below, we outline their formulations.
The $\mathtt{NICE^1}$ metric is defined as:
\begin{equation}
\mathtt{NICE^1} = \frac{\mathtt{MAE_X}}{\mathtt{MAE_P}},
\end{equation}
where $\mathtt{MAE_X}$ is the mean absolute error of the model, and $\mathtt{MAE_P}$ is the mean absolute error of the persistence model.
For a stationary time series, the differences $D = y(t) - y(t-\Delta t)$ are assumed to follow a normal distribution $\mathcal{N}(0, \sigma_D^2)$, where $\sigma_D^2 = 2 \sigma^2 (1 - \rho(1))$. The expected absolute value of a normal random variable is given by $\mathbb{E}[|D|] = \sqrt{\frac{2}{\pi}}   \sigma_D$. Substituting $\sigma_D$, the persistence error $\mathtt{MAE_P}$ is approximated as:
\begin{equation}
\mathtt{MAE_P} \approx \sqrt{\frac{2}{\pi}}   \sqrt{2   \sigma^2   (1 - \rho(1))}.
\end{equation}
Thus, the theoretical $\mathtt{NICE^1}$ becomes:
\begin{equation}
\mathtt{NICE^1} \approx \frac{\mathtt{MAE_X}}{\sqrt{\frac{2}{\pi}}   \sigma   \sqrt{2   (1 - \rho(1))}}.
\end{equation}
The $\mathtt{NICE^3}$ metric is defined as:
\begin{equation}
\mathtt{NICE^3} = \frac{\mathtt{RMCE_X}}{\mathtt{RMCE_P}},
\end{equation}
where $\mathtt{RMCE_X}$ is the root mean cubic error of the model while $\mathtt{RMCE_P}$ is focused about the persistence model.
The expected third absolute moment of a normal random variable $D$ is given by:
\begin{equation}
\mathbb{E}[|D|^3] = \frac{4\sqrt{2}}{\sqrt{\pi}}   \sigma_D^3.
\end{equation}
Substituting $\sigma_D = \sqrt{2   \sigma^2   (1 - \rho(1))}$, the persistence error $\mathtt{RMCE_P}$ becomes:
\begin{equation}
\mathtt{RMCE_P} \approx \left(\frac{4\sqrt{2}}{\sqrt{\pi}}\right)^{1/3}   \sigma   \sqrt{2   (1 - \rho(1))}.
\end{equation}
Thus, the theoretical $\mathtt{NICE^3}$ is expressed as:
\begin{equation}
\mathtt{NICE^3} \approx \frac{\mathtt{RMCE_X}}{\left(\frac{4\sqrt{2}}{\sqrt{\pi}}\right)^{1/3}   \sigma   \sqrt{2   (1 - \rho(1))}}.
\end{equation}
The derivation of $\mathtt{NICE^1}$ and $\mathtt{NICE^3}$ relies on several key assumptions. First, the time series $y(t)$ is assumed to be weakly stationary, meaning the variance $\sigma^2$ and autocorrelation $\rho(1)$ remain constant over time. Second, the differences $D = y(t) - y(t-\Delta t)$ are assumed to follow a normal distribution $\mathcal{N}(0, \sigma_D^2)$, where $\sigma_D^2 = 2\sigma^2 (1 - \rho(1))$. Third, the sample mean absolute error ($\mathtt{MAE_P}$) and mean cubic error ($\mathtt{RMCE_P}$) are approximated by their expected values ($\mathbb{E}[|D|]$ and $\mathbb{E}[|D|^3]$).
In practice, deviations from the theoretical values may occur due to several factors. Non-normality of differences can arise because experimental time series often exhibit non-Gaussian noise or extreme outliers. Non-stationarity, including variations in $\sigma^2$ or $\rho(1)$ over time, can lead to inconsistencies in the theoretical estimates. Additionally, finite sample effects can introduce biases in the empirical estimates of $\sigma$ and $\rho(1)$, particularly when the sample size is small. 
The presented formulas provide a robust theoretical framework for evaluating forecasting accuracy, but care should be taken when applying them to non-ideal datasets.

\section{Limitations of The Theoretical Approach}
\label{annex:limitation}
$\mathtt{NICE^k}$, despite its utility, comes with several limitations. One key assumption is stationarity, which underpins the link between the persistence model’s error and the lag-1 autocorrelation coefficient, $\rho(1)$. Stationarity implies that the variance and covariance of the series remain constant over time, which is only approximately valid for many time series, such as $\mathtt{I}$. However, in practical contexts, a weaker assumption of local stationarity at lag 1 often holds, especially for time series with strong short-term autocorrelation or periodicity. This makes the metric a reasonable approximation even for some non-stationary signals.
Additionally, some theoretical derivations of $\mathtt{NICE^k}$, particularly for $\mathtt{k} = 1$ and $\mathtt{k} = 3$, rely on an assumption of normality to compute expectation values such as $\mathbb{E}[|y(t) - y(t-\Delta t)|]$ and $\mathbb{E}[(y(t) - y(t-\Delta t))^3]$. In reality, many time series exhibit skewed or heavy-tailed distributions, particularly in the presence of extreme events. When the normality assumption does not hold, deviations in empirical estimates of $\mathtt{NICE^k}$ may occur, requiring cautious interpretation.
Moreover, in strongly non-stationary time series where variance exhibits significant drifts over time, the normalization process in $\mathtt{NICE^k}$ may become unstable. If the reference error (e.g., $\mathtt{RMSE_P}$) varies significantly across different periods, the metric may provide inconsistent comparisons between different forecasting models, particularly for long-term predictions.
Another limitation of $\mathtt{nRMSE_P}$ is its sensitivity to outliers. As a squared-error-based metric, the $\mathtt{RMSE}$ is disproportionately affected by extreme values, which can distort the evaluation of forecasting performance. For time series that exhibit occasional large deviations, such as those driven by extreme weather conditions, this sensitivity may lead to overestimation of forecasting errors. In such cases, alternative metrics, like the normalized Mean Absolute Error ($\mathtt{nMAE}$), which are less influenced by outliers, might provide a more robust evaluation of forecasting performance.
Finally, the metric assumes a perfect predictor as its lower bound, even though this is unattainable in practice. Time series inherently contain components that are unpredictable due to stochastic processes or unknown external factors. While complex forecasting approaches can reduce error significantly, they cannot fully eliminate the portion of variability that is fundamentally non-predictable. 
Despite these limitations, $\mathtt{NICE^k}$ remains a valuable tool for benchmarking forecasting models across different datasets. Its bounded and normalized nature enables interpretable comparisons, but its assumptions should always be considered when applying it to complex or highly non-stationary time series. When necessary, hybrid evaluation approaches combining multiple error metrics can help mitigate these challenges and provide a more comprehensive assessment of forecasting performance. The empirical method (see section \ref{empirical}) for calculating $\mathtt{NICE^k}$ is certainly the best solution, since no assumptions are required.

\section{Interpretation and Physical Sense}
\label{annex:interpretation}
The $\mathtt{L^k}\mathtt{-Error_P}$ in general and $\mathtt{nRMSE_{P}}$ in particular, provide a normalized measure of forecasting accuracy relative to the persistence model, the simplest and most basic forecasting approach where $\hat{y}(t) = y(t-n\Delta t)$. Persistence is computationally trivial but often the least effective model, as it ignores dynamic factors and external influences affecting time series behavior. In fields such as solar radiation ($\mathtt{I}$), persistence is frequently used as a baseline since it represents the minimum forecasting effort. However, it is well-documented that persistence struggles to predict rapid changes or non-periodic variations, making it a useful but simplistic reference point.
The $\mathtt{NICE^k}$ metric is interpreted on a bounded scale between 0 and 1 for practical use. A value close to 0 indicates that the forecasting model significantly outperforms persistence, capturing more of the underlying dynamics of the time series. A value of 1 corresponds to performance equal to persistence, which is undesirable but provides a benchmark for minimally acceptable forecasting. Values greater than 1 indicate that the model performs worse than persistence, failing to capture even basic temporal dependencies within the series.
Beyond its direct interpretation as a normalized error metric, $\mathtt{NICE^k}$ can also be linked to the statistical properties of the time series, particularly the $\mathtt{k}$-order moments of consecutive differences, $y(t) - y(t-\Delta t)$. This connection provides deeper insight into the role of temporal variability in forecasting accuracy. In the context of $\mathtt{I}$, and considering the clear sky index $\mathtt{K_c}$ as a stationary process defined by:
\begin{equation}
    \mathtt{K_c}(t) = \frac{\mathtt{I}(t)}{\mathtt{I}_{\mathtt{clr}}(t)},
\end{equation}
where $\mathtt{I}_{\mathtt{clr}}(t)$ is the clear sky irradiance. The variance (second-order moment) of $\Delta \mathtt{K_c} = \mathtt{K_c}(t) - \mathtt{K_c}(t-\Delta t)$ is expressed as:
\begin{equation}
    \sigma^2(\Delta \mathtt{K_c}) = 2\sigma^2(\mathtt{K_c})   (1 - \rho(1)).
\end{equation}
This is a standard approach for estimating the variability of $\mathtt{I}$ \cite{HOFF20101782}. 
While simple persistence assumes that future values remain unchanged, it fails to account for systematic diurnal variations in solar radiation. To address this, the Smart Persistence ($\mathtt{SP}$) model provides an improved baseline by incorporating predictable daily patterns, using the clear sky index to adjust forecasts. Unlike the naive persistence model, which assumes that future values remain constant, the smart persistence model adjusts predictions using the clear sky index, thus accounting for systematic daily variability:
\begin{equation}
    \mathtt{\hat{I}}(t+n\Delta t)=\mathtt{I}(t)\mathtt{I_{clr}}(t+n\Delta t)/\mathtt{I_{clr}}(t).
\end{equation}
Expanding the relationship further, the $\mathtt{RMSE}$ of the smart persistence model can be expressed as:
\begin{equation}
    \mathtt{RMSE_{SP}} = \sqrt{2}   \sigma(\mathtt{K_c})   \sqrt{1 - \rho(1)}   \sqrt{\sigma^2(\mathtt{I}_{\mathtt{clr}}) + \mu_{\mathtt{clr}}^2}=\sigma(\Delta \mathtt{K_c}) \mu_{\mathtt{clr}}\sqrt{1+\frac{\sigma^2_{\mathtt{clr}}}{\mu^2_{\mathtt{clr}}}}.
\end{equation}
This formulation explicitly shows how $\mathtt{RMSE_{SP}}$ depends on the variability of $\mathtt{K_c}$ and the characteristics of clear sky irradiance. The first term, $\sigma(\Delta \mathtt{K_c})$, captures the short-term fluctuations of the clear sky index, while $\mu_{\mathtt{clr}}$ and $\sigma_{\mathtt{clr}}$ account for the mean and variance of clear sky irradiance. This highlights why $\mathtt{SP}$ provides a refined baseline for solar forecasting, as it adjusts for systematic variations in $\mathtt{I}_{\mathtt{clr}}$. 
Assuming $\Delta \mathtt{K_c}$ and $\mathtt{I_{clr}(t)}$ are uncorrelated, the $\mathtt{RMSE}$ of the smart persistence model predicting $\mathtt{I}$ is given, after some manipulations, by:
\begin{equation}
    \mathtt{RMSE_{SP}} \propto \sigma(\Delta \mathtt{K_c}).
\end{equation}
This relationship highlights the impact of short-term variability in both $\mathtt{K_c}$ and $\mathtt{I}_{\mathtt{clr}}$ on forecasting accuracy.
It is crucial to define error metrics that are independent of inherent variability, ensuring that errors remain unbiased and do not merely reflect fluctuations in the underlying time series. Using simple persistence instead of Smart Persistence when computing $\mathtt{NICE^k}$ ensures that the metric remains independent of systematic variations, focusing solely on the forecasting model's ability to capture temporal dependencies beyond trivial baselines.
The $\mathtt{NICE^2}$ framework encapsulates this trade-off by setting an idealized lower bound at zero, even though perfect predictions remain theoretical. This ensures that $\mathtt{NICE^2}$ provides a meaningful and interpretable scale for assessing the relative accuracy of forecasting models. By incorporating a robust normalization strategy, it enables fair comparisons across different forecasting approaches and applications, making it a valuable tool for evaluating predictive performance in time series forecasting.

\bibliographystyle{elsarticle-num} 
\bibliography{Bib} 

\begin{thebibliography}{10}
\expandafter\ifx\csname url\endcsname\relax
  \def\url#1{\texttt{#1}}\fi
\expandafter\ifx\csname urlprefix\endcsname\relax\def\urlprefix{URL }\fi
\expandafter\ifx\csname href\endcsname\relax
  \def\href#1#2{#2} \def\path#1{#1}\fi

\bibitem{lauret2012bayesianmodelcommitteeapproach}
P.~Lauret, A.~Rodler, M.~Muselli, M.~David, H.~Diagne, C.~Voyant, A {Bayesian}
  model committee approach to forecasting global solar radiation (2012).
\newblock \href {http://arxiv.org/abs/1203.5446} {\path{arXiv:1203.5446}}.

\bibitem{9279099}
G.~Notton, M.-L. Nivet, C.~Voyant, J.-L. Duchaud, A.~Fouilloy, D.~Zafirakis,
  J.~Kaldellis, {Tilos, an autonomous Greek island thanks to a PV/Wind/Zebra
  battery plant and a smart Energy Management System}, in: 2020 7th
  International Conference on Energy Efficiency and Agricultural Engineering
  (EE\&AE), 2020, pp. 1--4.
\newblock \href {https://doi.org/10.1109/EEAE49144.2020.9279099}
  {\path{doi:10.1109/EEAE49144.2020.9279099}}.

\bibitem{manjili2018data}
Y.~S. Manjili, R.~Vega, M.~Jamshidi, Data-analytic-based adaptive solar energy
  forecasting framework, IEEE Systems Journal 12 (2018) 285--296.
\newblock \href {https://doi.org/10.1109/JSYST.2017.2769483}
  {\path{doi:10.1109/JSYST.2017.2769483}}.

\bibitem{voyant2017machine}
C.~Voyant, G.~Notton, S.~Kalogirou, M.~Nivet, C.~Paoli, F.~Motte, A.~Fouilloy,
  Machine learning methods for solar radiation forecasting: A review, Renewable
  Energy 105 (2017) 569--582.
\newblock \href {https://doi.org/10.1016/J.RENENE.2016.12.095}
  {\path{doi:10.1016/J.RENENE.2016.12.095}}.

\bibitem{bacher2009short}
P.~Bacher, H.~Madsen, H.~A. Nielsen, Short-term solar power forecasting—an
  evaluation of different forecasting methods, Solar Energy 83~(9) (2009)
  1772--1783.
\newblock \href {https://doi.org/10.1016/j.solener.2009.05.016}
  {\path{doi:10.1016/j.solener.2009.05.016}}.

\bibitem{yang2014solar}
R.~Yang, B.~Kurtz, J.~Kleissl, Solar irradiance forecasting using a
  ground-based sky imager developed at {UC San Diego}, Solar Energy 103 (2014)
  502--524.
\newblock \href {https://doi.org/10.1016/j.solener.2014.02.044}
  {\path{doi:10.1016/j.solener.2014.02.044}}.

\bibitem{taylor2016forecasting}
C.~Vennila, A.~Titus, T.~S. Sudha, U.~Sreenivasulu, N.~P.~R. Reddy, K.~Jamal,
  D.~Lakshmaiah, P.~Jagadeesh, A.~Belay, Forecasting solar energy production
  using machine learning, International Journal of Photoenergy 2022~(1) (2022)
  7797488.
\newblock \href {https://doi.org/10.1155/2022/7797488}
  {\path{doi:10.1155/2022/7797488}}.

\bibitem{yang2020verification}
D.~Yang, S.~Alessandrini, J.~Antonanzas, F.~Antonanzas-Torres, V.~Badescu,
  H.~G. Beyer, R.~Blaga, J.~Boland, J.~M. Bright, C.~F. Coimbra, M.~David,
  Âzeddine Frimane, C.~A. Gueymard, T.~Hong, M.~J. Kay, S.~Killinger,
  J.~Kleissl, P.~Lauret, E.~Lorenz, D.~{van der Meer}, M.~Paulescu, R.~Perez,
  O.~Perpiñán-Lamigueiro, I.~M. Peters, G.~Reikard, D.~Renné, Y.-M.
  Saint-Drenan, Y.~Shuai, R.~Urraca, H.~Verbois, F.~Vignola, C.~Voyant,
  J.~Zhang, Verification of deterministic solar forecasts, Solar Energy 210
  (2020) 20--37, special Issue on Grid Integration.
\newblock \href {https://doi.org/10.1016/j.solener.2020.04.019}
  {\path{doi:10.1016/j.solener.2020.04.019}}.

\bibitem{zhang2015suite}
J.~Zhang, A.~Florita, B.~Hodge, S.~Lu, H.~Hamann, V.~Banunarayanan, A.~M.
  Brockway, A suite of metrics for assessing the performance of solar power
  forecasting, Solar Energy 111 (2015) 157--175.
\newblock \href {https://doi.org/10.1016/J.SOLENER.2014.10.016}
  {\path{doi:10.1016/J.SOLENER.2014.10.016}}.

\bibitem{hansen2019arbiter}
C.~W. Hansen, W.~Holmgren, A.~e.~a. Tuohy, The solar forecast arbiter: An open
  source evaluation framework for solar forecasting, in: 2019 IEEE 46th
  Photovoltaic Specialists Conference (PVSC), 2019, pp. 2452--2457.
\newblock \href {https://doi.org/10.1109/PVSC40753.2019.8980713}
  {\path{doi:10.1109/PVSC40753.2019.8980713}}.

\bibitem{vallance2017standardized}
L.~Vallance, B.~Charbonnier, N.~Paul, S.~Dubost, P.~Blanc, Towards a
  standardized procedure to assess solar forecast accuracy: A new ramp and time
  alignment metric, Solar Energy 150 (2017) 408--422.
\newblock \href {https://doi.org/10.1016/J.SOLENER.2017.04.064}
  {\path{doi:10.1016/J.SOLENER.2017.04.064}}.

\bibitem{marquez2013proposed}
R.~Marquez, C.~F. Coimbra, Proposed metrics for solar forecasting model
  evaluation, Journal of Solar Energy Engineering 135~(1) (2013) 011016.
\newblock \href {https://doi.org/10.1115/1.4007496}
  {\path{doi:10.1115/1.4007496}}.

\bibitem{willmott2005advantages}
C.~J. Willmott, K.~Matsuura, {Advantages of the mean absolute error (MAE) over
  the root mean square error (RMSE) in assessing average model performance},
  Climate Research 30~(1) (2005) 79--82.
\newblock \href {https://doi.org/10.3354/cr030079}
  {\path{doi:10.3354/cr030079}}.

\bibitem{SS}
A.~H. Murphy, Skill scores based on the mean square error and their
  relationships to the correlation coefficient, Monthly Weather Review 116~(12)
  (1988) 2417 -- 2424.
\newblock \href
  {https://doi.org/10.1175/1520-0493(1988)116<2417:SSBOTM>2.0.CO;2}
  {\path{doi:10.1175/1520-0493(1988)116<2417:SSBOTM>2.0.CO;2}}.

\bibitem{perez2017value}
O.~Gandhi, W.~Zhang, D.~S. Kumar, C.~D. Rodríguez-Gallegos, G.~M. Yagli,
  D.~Yang, T.~Reindl, D.~Srinivasan, The value of solar forecasts and the cost
  of their errors: A review, Renewable and Sustainable Energy Reviews 189
  (2024) 113915.
\newblock \href {https://doi.org/10.1016/j.rser.2023.113915}
  {\path{doi:10.1016/j.rser.2023.113915}}.

\bibitem{lorenz2009forecast}
E.~Lorenz, D.~Heinemann, H.~Wickramarathne, H.~Beyer, S.~Bofinger, Forecast of
  ensemble power production by grid-connected {PV} systems, 20th European
  Photovoltaic Solar Energy Conference, 2007, pp. 3--9.

\bibitem{YANG2019410}
D.~Yang, A universal benchmarking method for probabilistic solar irradiance
  forecasting, Solar Energy 184 (2019) 410--416.
\newblock \href {https://doi.org/j.solener.2019.04.018}
  {\path{doi:j.solener.2019.04.018}}.

\bibitem{foley2012current}
A.~M. Foley, P.~G. Leahy, A.~Marvuglia, E.~J. McKeogh, Current methods and
  advances in forecasting of wind power generation, Renewable Energy 37~(1)
  (2012) 1--8.
\newblock \href {https://doi.org/10.1016/j.renene.2011.05.033}
  {\path{doi:10.1016/j.renene.2011.05.033}}.

\bibitem{Singla2021}
P.~Singla, M.~Duhan, S.~Saroha, Review of different error metrics: A case of
  solar forecasting, AIUB Journal of Science and Engineering 20~(4) (2021)
  158--165.
\newblock \href {https://doi.org/10.53799/ajse.v20i4.212}
  {\path{doi:10.53799/ajse.v20i4.212}}.

\bibitem{10.1063/5.0042710}
C.~Voyant, P.~Lauret, G.~Notton, J.-L. Duchaud, A.~Fouilloy, M.~David, Z.~M.
  Yaseen, T.~Soubdhan, A {Monte Carlo} based solar radiation forecastability
  estimation, Journal of Renewable and Sustainable Energy 13~(2) (2021) 026501.
\newblock \href {https://doi.org/10.1063/5.0042710}
  {\path{doi:10.1063/5.0042710}}.

\bibitem{marquez2013metric}
R.~Marquez, C.~Coimbra, Proposed metric for evaluation of solar forecasting
  models, Journal of Solar Energy Engineering 135 (2013) 011016.
\newblock \href {https://doi.org/10.1115/1.4007496}
  {\path{doi:10.1115/1.4007496}}.

\bibitem{ttt}
M.~P. Mittermaier, A “meta” analysis of the fractions skill score: The
  limiting case and implications for aggregation, Monthly Weather Review
  149~(10) (2021) 3491 -- 3504.
\newblock \href {https://doi.org/10.1175/MWR-D-18-0106.1}
  {\path{doi:10.1175/MWR-D-18-0106.1}}.

\bibitem{Nguyen2022AMO}
T.~N.~A. Nguyen, F.~M{\"u}sgens, A meta-analysis of solar forecasting based on
  skill score, ArXiv abs/2208.10536 (2022).

\bibitem{VOYANT2022747}
C.~Voyant, G.~Notton, J.-L. Duchaud, L.~A.~G. Gutiérrez, J.~M. Bright,
  D.~Yang, Benchmarks for solar radiation time series forecasting, Renewable
  Energy 191 (2022) 747--762.
\newblock \href {https://doi.org/10.1016/j.renene.2022.04.065}
  {\path{doi:10.1016/j.renene.2022.04.065}}.

\bibitem{VOYANT2026123913}
C.~Voyant, A.~Julien, M.~Despotovic, G.~Notton, L.~A. Garcia-Gutierrez, C.~F.
  Nicolosi, P.~Blanc, J.~Bright, Stochastic coefficient of variation: Assessing
  the variability and forecastability of solar irradiance, Renewable Energy 256
  (2026) 123913.
\newblock \href {https://doi.org/10.1016/j.renene.2025.123913}
  {\path{doi:10.1016/j.renene.2025.123913}}.

\bibitem{despotovic2024}
M.~Despotovic, C.~Voyant, L.~Garcia-Gutierrez, J.~Almorox, G.~Notton, Solar
  irradiance time series forecasting using auto-regressive and extreme learning
  methods: Influence of transfer learning and clustering, Applied Energy 365
  (2024) 123215.
\newblock \href {https://doi.org/10.1016/j.apenergy.2024.123215}
  {\path{doi:10.1016/j.apenergy.2024.123215}}.

\bibitem{VOYANT2025113490}
C.~Voyant, M.~Despotovic, G.~Notton, Y.-M. Saint-Drenan, M.~Asloune,
  L.~Garcia-Gutierrez, On the importance of clearsky model in short-term solar
  radiation forecasting, Solar Energy 294 (2025) 113490.
\newblock \href {https://doi.org/10.1016/j.solener.2025.113490}
  {\path{doi:10.1016/j.solener.2025.113490}}.

\bibitem{forstinger:hal-03657585}
A.~Forstinger, S.~Wilbert, A.~Jensen, B.~Kraas, C.~Fern{\'a}ndez~Peruchena,
  C.~Gueymard, D.~Ronzio, D.~Yang, E.~Collino, J.~Polo~Martinez, J.~Ruiz-Arias,
  N.~Hanrieder, P.~Blanc, Y.-M. Saint-Drenan,
  \href{https://minesparis-psl.hal.science/hal-03657585}{{Expert quality
  control of solar radiation ground data sets}}, in: {ISES Solar World
  Congress}, online, France, 2021.
\newline\urlprefix\url{https://minesparis-psl.hal.science/hal-03657585}

\bibitem{jarque1987test}
C.~M. Jarque, A.~K. Bera, A test for normality of observations and regression
  residuals, International Statistical Review / Revue Internationale de
  Statistique 55~(2) (1987) 163--172.
\newblock \href {https://doi.org/10.2307/1403192} {\path{doi:10.2307/1403192}}.

\bibitem{PhysRevX.13.021018}
F.~Mogavero, N.~H. Hoang, J.~Laskar, Timescales of chaos in the inner solar
  system: Lyapunov spectrum and quasi-integrals of motion, Phys. Rev. X 13
  (2023) 021018.
\newblock \href {https://doi.org/10.1103/PhysRevX.13.021018}
  {\path{doi:10.1103/PhysRevX.13.021018}}.

\bibitem{HOFF20101782}
T.~E. Hoff, R.~Perez, Quantifying {PV} power output variability, Solar Energy
  84~(10) (2010) 1782--1793.
\newblock \href {https://doi.org/10.1016/j.solener.2010.07.003}
  {\path{doi:10.1016/j.solener.2010.07.003}}.

\end{thebibliography}

\end{document}